\documentclass[12pt]{article}

\pdfoutput=1
\usepackage[totalheight=23cm, totalwidth=17cm]{geometry}
\usepackage{amssymb,amsmath,graphicx,subfigure,color,cite,bm,ulem}
\usepackage[colorlinks=true,linkcolor=blue,citecolor=blue]{hyperref}

\begin{document}

\begin{titlepage}

\vskip 2cm

\begin{center}

{\Large \bf
Distribution of primordial black holes \\[2mm] and 21cm signature
}

\vspace{1cm}

Jinn-Ouk Gong$^{\,a}$
and
Naoya Kitajima$^{\,b}$ \\

\vskip 1.0cm

{\it
$^a$Korea Astronomy and Space Science Institute, Daejeon 34055, Korea
\\[2mm]
$^b$Department of Physics, Nagoya University, Nagoya 464-8602, Japan\\
}

\vskip 1.0cm

\begin{abstract}

We show that the number of primordial black holes (PBHs) which is originated from primordial density perturbations with moderately-tilted power spectrum fluctuates following the log-normal distribution, while it follows the Poisson distribution if the spectrum is steeply blue. The log-normal, as well as the Poisson, fluctuation of the PBH number behaves as an isocurvature mode and affects the matter power spectrum and the halo mass function in a different way from those for the Poisson case. The future 21cm observation can potentially put a stronger constraint on the PBH fraction than the current one in a wide mass range, $10^{-5}M_\odot$--$10M_\odot$.

\end{abstract}

\end{center}

\end{titlepage}

\newpage

\section{Introduction} 
\label{sec:intro}

The observations of binary black hole merger events by the direct detection of gravitational waves \cite{Abbott:2016blz,Abbott:2016nmj,Abbott:2017vtc,Abbott:2017gyy,Abbott:2017oio} can give us hints to investigate the history of the early Universe. In particular, one can explain those events by primordial black holes (PBHs) \cite{Bird:2016dcv,Clesse:2016vqa,Sasaki:2016jop}. PBH is the black hole originated from the direct collapse of cosmic fluids in the early Universe \cite{Zeldovich,Hawking:1971ei,Carr:1974nx}. The formation of PBHs needs large initial density perturbations on horizon scales which are typically seeded by inflationary fluctuations.

PBHs can contribute to the present dark matter component if their mass is heavy enough to survive  the age of the Universe without being evaporated away. If they exist, they leave various cosmological and astrophysical imprints such as the gravitational microlensing \cite{Allsman:2000kg,Tisserand:2006zx,Niikura:2017zjd}, the halo wide binaries \cite{Yoo:2003fr}, the cosmic microwave background (CMB) \cite{Ricotti:2007au,Chen:2016pud,Ali-Haimoud:2016mbv,Horowitz:2016lib} and the stochastic gravitational wave background \cite{Raidal:2017mfl}, see \cite{Carr:2009jm,Carr:2016drx,Carr:2017jsz,Sasaki:2018dmp,Carr:2018rid} for recent reviews.
In addition, PBHs have an unique feature coming from their formation process. Because PBH formation is a rare event and PBHs are sparsely distributed in space, there is a Poisson noise on the number of PBHs \cite{Meszaros:1975ef}. It behaves as an isocurvature mode and affects the structure formation on small scales. Then, PBHs can be constrained by Ly-$\alpha$ observation \cite{Afshordi:2003zb} and CMB measurement \cite{Chisholm:2005vm}, and can explain the observation of cosmic infrared background \cite{Kashlinsky:2016sdv}.

Because PBHs can exist well before the star formation epoch as opposed to stellar black holes, they can leave cosmological imprints in the Universe before the star formation, or ``dark age''. The measurement of the redshifted 21cm emission/absorption signal by radio telescope array such as Square Kilometre Array (SKA) \cite{SKA} is one of most promising future observations for the dark age. Then, the 21cm observations have the potential to detect or constrain PBHs. In particular, PBHs with large masses can affect the 21cm signature through matter accretion and emission of extra energetic photons \cite{Tashiro:2012qe,Bernal:2017nec}. Evaporating PBHs in the dark age can also change the reionization history by heating nearby intergalactic medium through Hawking radiation, which affects the 21cm signature \cite{Mack:2008nv}.

The sky-averaged signal of the 21cm fluctuation is sensitive to the number of minihaloes \cite{Iliev:2002gj}. In particular, blue-tilted isocurvature perturbations can significantly modify the number of minihaloes and enhance the 21cm emission signals \cite{Takeuchi:2013hza,Sekiguchi:2013lma}.
Our previous work \cite{Gong:2017sie} focuses on the modification of the halo mass function by isocurvature perturbations originated from the Poisson noise in the number of PBHs and discusses the possibility to constrain the PBH abundance by future 21cm survey. This concludes that the future SKA-like observations can constrain the PBH abundance with mass $M_{\rm PBH}>10^{-2}M_\odot$. It can be stronger than the current constraint and can probe those PBHs explaining the LIGO event.

However, in \cite{Gong:2017sie}, in order to ensure that the PBH number follows the Poisson distribution, we made an assumption that the PBH formation event in each horizon patch is completely independent and randomly occurs. This assumption is valid if PBHs are originated from sub-horizon physics such as the collapse of cosmic strings \cite{Polnarev:1988dh} or bubble collisions \cite{Hawking:1982ga}. 
However, it is doubtful if one assumes the inflationary fluctuations for the seeds of the PBH formation, 
in which super-horizon fluctuations are superimposed in each horizon patch. If so, the PBH formation event is no longer statistically independent in each horizon patch and the PBH number no longer follows the Poisson distribution function.

In this article, we study the distribution of the PBH number fluctuation in the presence of super-horizon perturbations at PBH formation, which is a typical consequence of inflation. Specifically, we adopt a locally-tilted power spectrum for the initial fluctuations and discuss how the distribution depends on the local spectral index of the density contrast, $n_\delta$, and show that it follows the Poisson distribution for highly blue-tilted spectrum, $n_\delta \gtrsim 3$, and for moderately-tilted case with $n_\delta \lesssim 2$, it follows the log-normal distribution function. 
The outline of this article is as follows.
We consider the distribution of the PBH number fluctuation both analytically and numerically and discuss the resultant isocurvature fluctuation and modification of the halo mass function in Section~\ref{sec:PBHfluc}. Then, we discuss the implications for the Ly-$\alpha$ constraint and the forecasted constraint by future 21cm survey in Section~\ref{sec:constraint}. We briefly conclude in Section~\ref{sec:disc}.

\section{PBH number fluctuations} 
\label{sec:PBHfluc}

\subsection{Distribution of PBH number}

Let us consider the following {\it locally} power-law dimensionless power spectrum of the initial density perturbations for the PBH formation:\footnote{
We start from the power spectrum of the {\it density contrast} instead of the curvature perturbation
because the curvature perturbation is not suitable to consider the effects from super-horizon modes \cite{Young:2014ana}. Note also that the spectral index $n_\delta$ in (\ref{eq:powerspec}) differs from that of the curvature perturbation, $n_s$, because the conversion formula between the density contrast and the (comoving) curvature perturbation has $k$-dependence.
}
\begin{equation} 
\label{eq:powerspec}
{\cal P}_\delta(k) = {\cal P}_\delta(k_*) \left( \frac{k}{k_*} \right)^{n_\delta-1}
\quad \text{for} \quad
k_c < k < k_* \, ,
\end{equation}
where we assume there is a cutoff wavenumber $k_*$ roughly corresponding to the scale of PBH formation and $k_c$ is some critical wavenumber below which the nearly scale-invariant spectrum dominates to be consistent with the CMB observations. 
One can consider the running of the spectral index, but here for simplicity we consider a constant spectral index within a narrow range of $k$ responsible for the PBH formation. 
Note that, in most cases, one needs $n_\delta > 1$ for PBH formation.
Such power spectrum can be realized in some classes of inflation model \cite{Kohri:2007qn,Drees:2011yz} or the curvaton scenario \cite{Kohri:2007qn,Kawasaki:2012wr}.

In the limit of steeply blue spectrum, the amplitude of the initial density perturbations increases rapidly so that the formation of PBHs at a certain length scale is hardly influenced by longer wavelength modes but is mostly determined by the value of $\delta$ with the corresponding wavelength. Consequently the PBH formation is random and independent event in each horizon patch, which leads to the fluctuation of the PBH number following the Poisson distribution function. However, it is not the case with relatively mild power spectrum with $n_\delta \sim 1$ because larger scale fluctuations are non-negligibly superposed when we smooth the density contrast over the PBH formation scale.

\subsubsection{PBH number counts in lattice space}

The PBH formation criterion is applied to the smoothed density contrast over a sphere with smoothing radius $R$ given by
\begin{equation} \label{eq:deltas}
\delta_s = \int d^3x' \delta (\bm{x}') W(|\bm{x}-\bm{x}'|,R) = \int \frac{d^3k}{(2\pi)^3} e^{i \bm{k} \cdot \bm{x}} W(kR) \delta(\bm{k}),
\end{equation}
where $\delta(\bm{x})$ and $W(|\bm{x}-\bm{x}'|)$ are the density contrast and the window function respectively and $\delta(\bm{k})$ and $W(kR)$ are their Fourier transformations.
Here we adopt the following Gaussian window function:
\begin{equation}
W(x,R) = \frac{e^{-x^2/(2R^2)}}{(2\pi R^2)^{3/2}} ~~~\text{and}~~~ W(kR) = e^{-k^2 R^2/2}.
\end{equation}
The variance of the smoothed density contrast can be calculated via
\begin{equation} \label{eq:sigma2R}
\sigma^2(R) = \int^{\infty}_0 \frac{dk}{k} \frac{k^3 P_\delta(k)}{2\pi^2} W^2(kR),
\end{equation}
where $P_\delta(k)$ is defined through
\begin{equation} \label{eq:Pk}
\langle \delta(\bm{k}) \delta(\bm{k}') \rangle = (2\pi)^2 \delta^{(3)}(\bm{k}+\bm{k}') P_\delta(k),
\end{equation}
with $\delta^{(3)}(\bm{k}+\bm{k}')$ being the Dirac $\delta$-function in 3-dimensional space.
Note that $P_\delta(k)$ and the dimensionless power spectrum, ${\cal P}_\delta(k)$ in Eq.\,(\ref{eq:powerspec}) is related as ${\cal P}_\delta(k) = k^3 P_\delta (k)/(2\pi^2)$ in the case with 3-dimensional space.

If PBHs are originated from primordial density perturbations, which are initially super-horizon scale, the formation occurs when the relevant scale reenters the horizon and the smoothed density contrast over that scale exceeds the critical value $\delta_{\rm cr} \approx 0.4$ \cite{Harada:2013epa}. Hence the number of PBHs or the PBH formation probability are roughly determined by the variance of the smoothed density contrast given by Eq.\,(\ref{eq:sigma2R}).

To see the spatial distribution of PBHs for a given power spectrum, we rely on the lattice space simulation.
We follow the procedure of the convolution picture \cite{Salmon1996,Pen:1997up,Bertschinger:2001ng} to generate a random Gaussian field as the density contrast at each grid point for given power spectra.
Specifically, we generate, at each grid point, uncorrelated random Gaussian field, $\xi(\bm{x})$, with zero mean and the variance $(N_{\rm grid}/L)^3$ with $N_{\rm grid}$ and $L$ respectively being the grid number and the length of the simulation box along a single axis.
Then, we get the density contrast satisfying Eq.\,(\ref{eq:Pk}) by the following convolution formula,
\begin{equation}
\delta(\bm{x}) = \int d^3 x' T(\bm{x}-\bm{x}') \xi(\bm{x}'),
\end{equation}
where $T(\bm{x})$ is the Fourier transformation of $T(k) = (P_\delta(k))^{1/2}$.
The smoothed density contrast (\ref{eq:deltas}) can also be obtained by inserting the window function in the convolution integral.
In our lattice calculation, we set maximum/minimum cutoff wave numbers above/below which the power spectrum is zero.
The maximum wave number, $k_{\rm max} = \pi N_{\rm grid}/L$, comes from the resolution of the lattice space and the minimum wave number, $k_{\rm min} = 2\pi/L$, comes from the finite volume of the simulation box.

\begin{figure}[tp]
\centering
\subfigure[$n_\delta = 3$]{
\includegraphics [width = 7.5cm, clip]{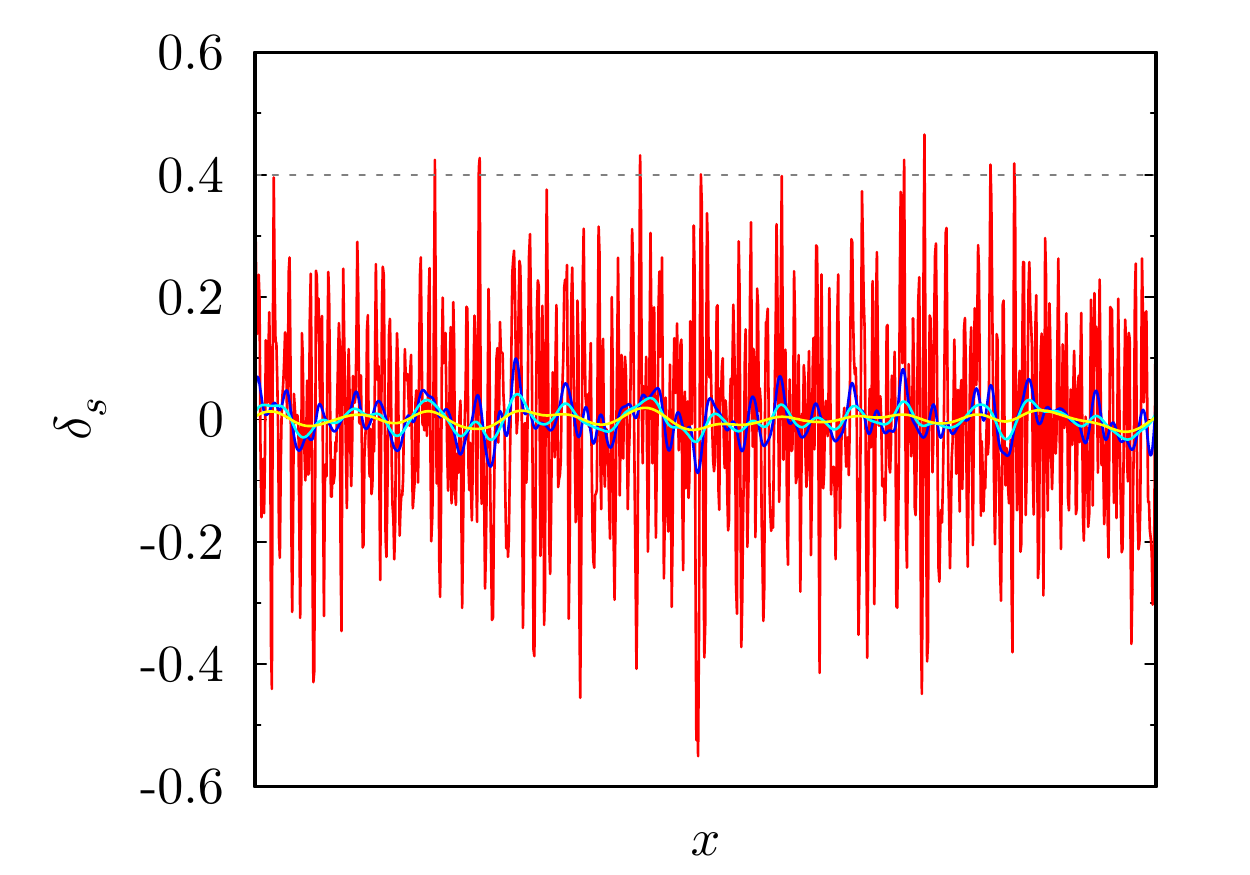}
\label{subfig:deltas1a}
}
\subfigure[$n_\delta = 1.5$]{
\includegraphics [width = 7.5cm, clip]{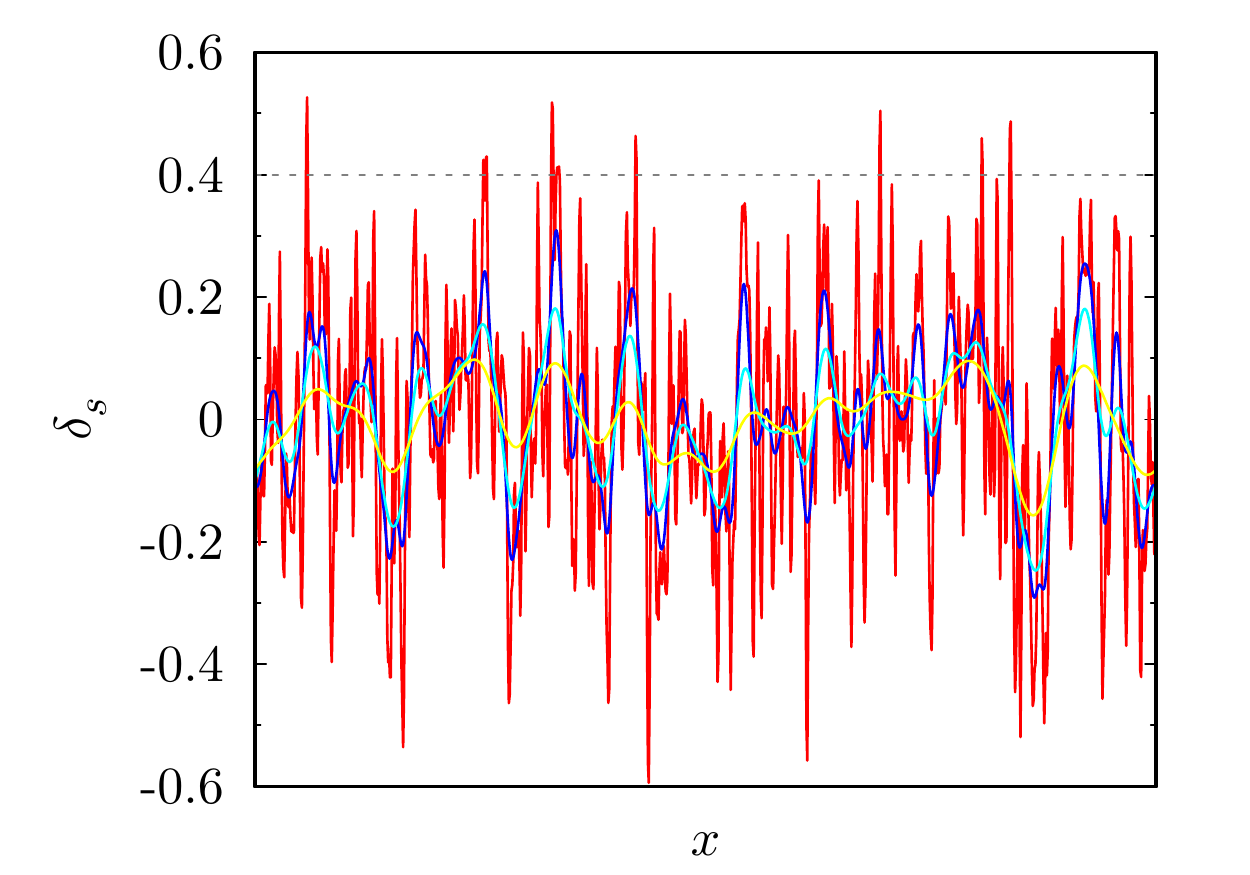}
\label{subfig:deltas1b}
}
\caption{
The smoothed density contrast in 1-dimensional lattice space with 1024 grids.
We have taken $n_\delta=3$ (left) and 1.5 (right).
}
\label{fig:deltas1}
\end{figure}

\begin{figure}[tp]
\centering
\subfigure[$n_\delta = 3$]{
\includegraphics [width = 7.0cm, clip]{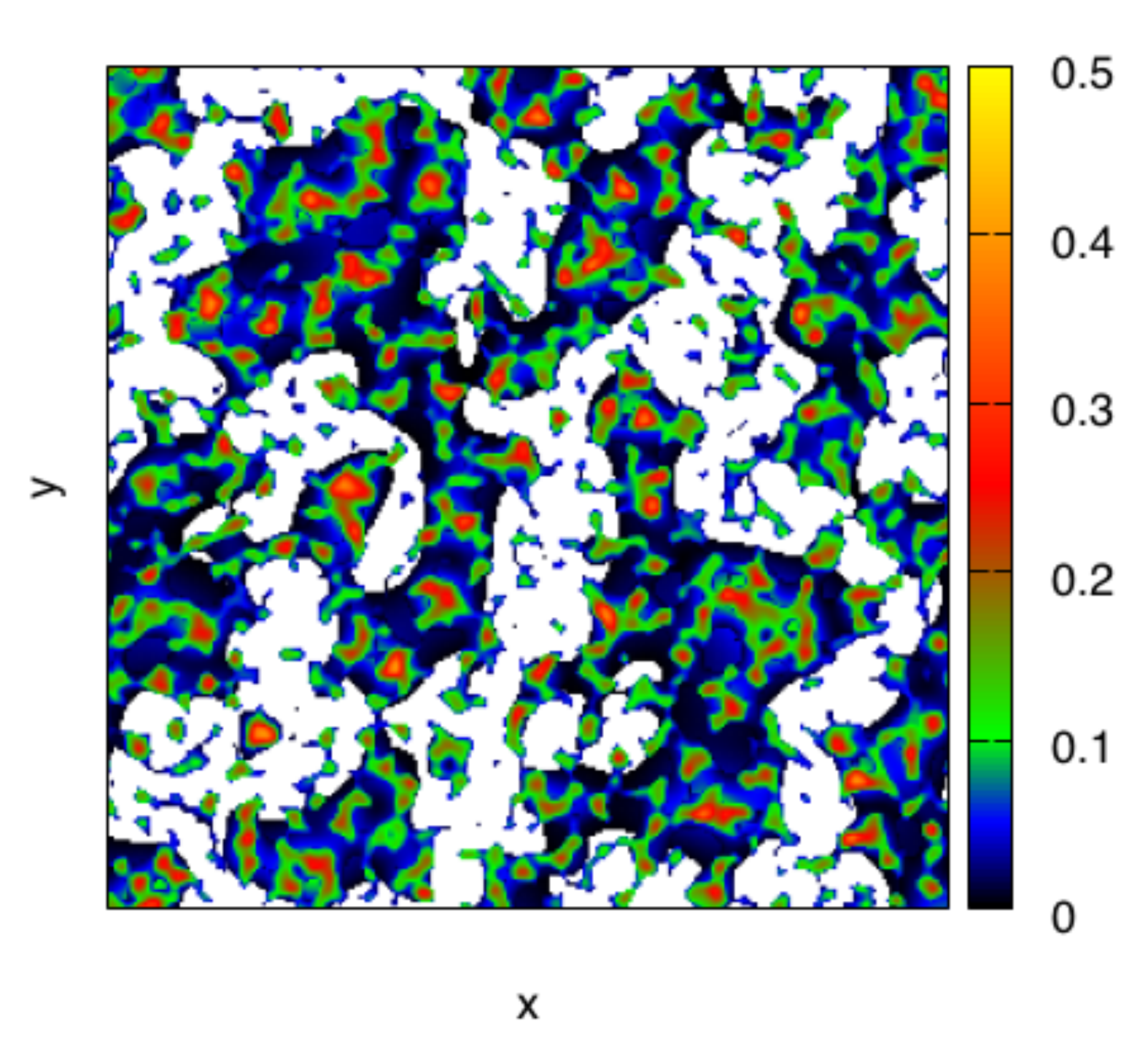}
\label{subfig:deltas2a}
}
\subfigure[$n_\delta = 1.5$]{
\includegraphics [width = 7.0cm, clip]{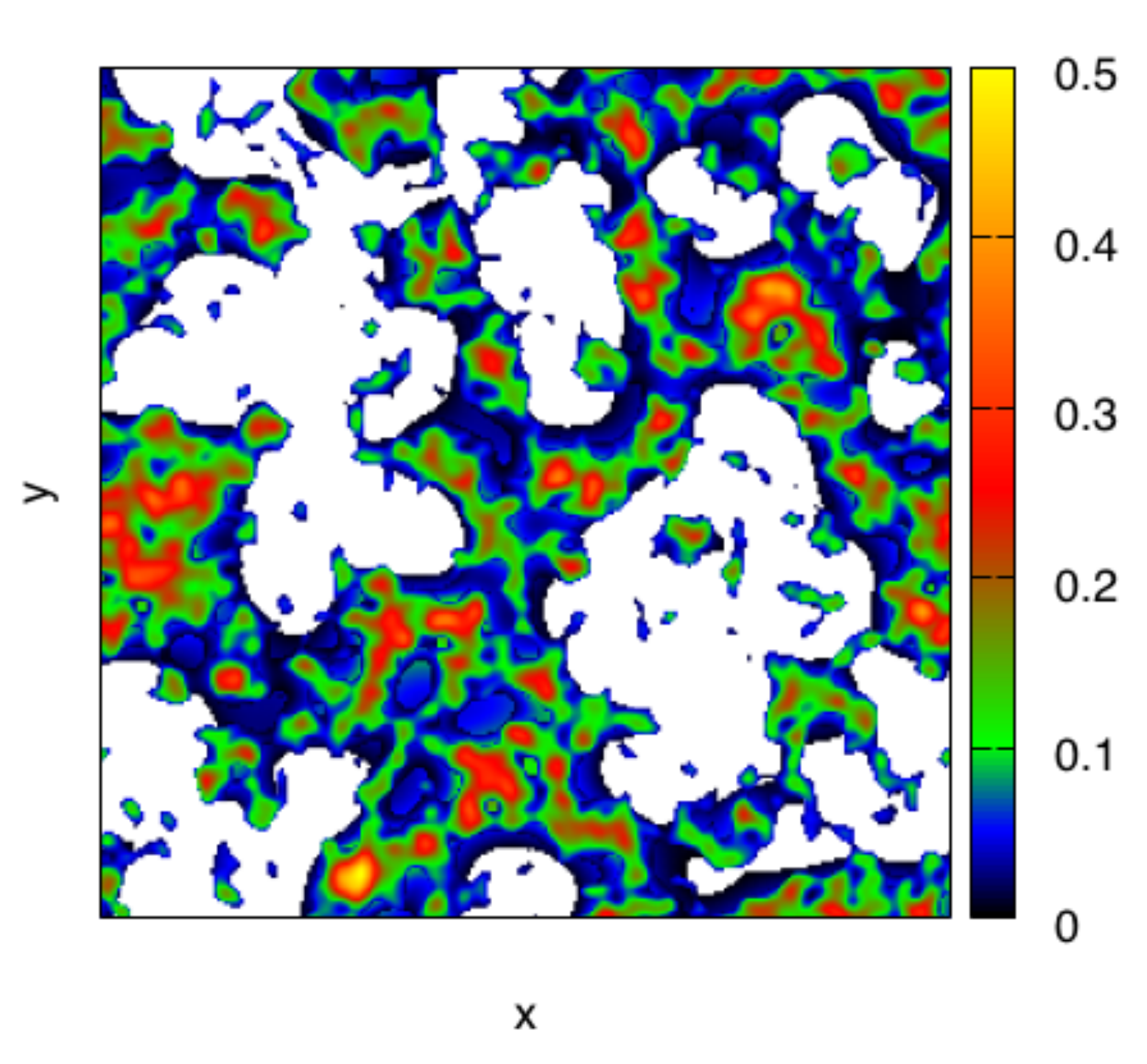}
\label{subfig:deltas2b}
}
\caption{
The smoothed density contrast in 2-dimensional lattice space with $512^2$ grids. The color bar shows the value of smoothed density contrast and the colorless region corresponds to underdensity region.
We have taken $n_\delta=3$ (left) and 1.5 (right).
}
\label{fig:deltas2}
\end{figure}

Fig.\,\ref{fig:deltas1} and \ref{fig:deltas2} show the smoothed density contrast in 1-dimensional and 2-dimensional lattice space respectively.
$\delta_s(R)$ with $R=L/N_{\rm grid}$, $5L/N_{\rm grid}$, $10L/N_{\rm grid}$ and $20L/N_{\rm grid}$ are superposed in each figure and we set $n_\delta=3$ (1.5) in each left (right) panel.
Note that the density contrast with $n_\delta = 1.5$ results in spatially more clustered configuration of overdensity regions, leading to spatially clustered PBHs.
Implications of such initial clustering of PBHs are studied in the literature \cite{Chisholm:2005vm,Tada:2015noa,Clesse:2016vqa,Desjacques:2018wuu}. Here we focus only on the fluctuation of the PBH number and quantitative study of such clustering property is beyond the scope of this paper. We left it for future work.

After generating a smoothed density field in each grid point in the lattice space, we count the region satisfying the PBH formation criterion.
Specifically, we identify a PBH as the region where the smoothed density contrast is a local peak and exceeds the critical value $\delta_{\rm cr} = 0.4$.\footnote{
A mildly-tilted density perturbation may yield uniformly dense regions over the scale larger than the PBH formation scale and it may not lead to the PBH formation even if the smoothed density contrast is locally larger than the critical value.
To discard such regions, we have also considered the {\it local} density contrast, $\delta\rho/\bar\rho_{\rm loc}$, whose denominator is not the global average but rather the local average over twice larger than the smoothing scale relevant to the PBH formation. However, we have found that the result does not change significantly. Then, we neglect such a issue throughout the paper.
}
The PBH number distribution can be obtained by generating a large number realizations and counting the number of PBHs in each realization. It should be noted that the variance of the smoothed density contrast directly obtained in the lattice space agrees well with the analytic formula (\ref{eq:sigma2R}).

\begin{figure}[tp]
\centering
\subfigure[$n_\delta = 1.5$]{
\includegraphics [width = 7.5cm, clip]{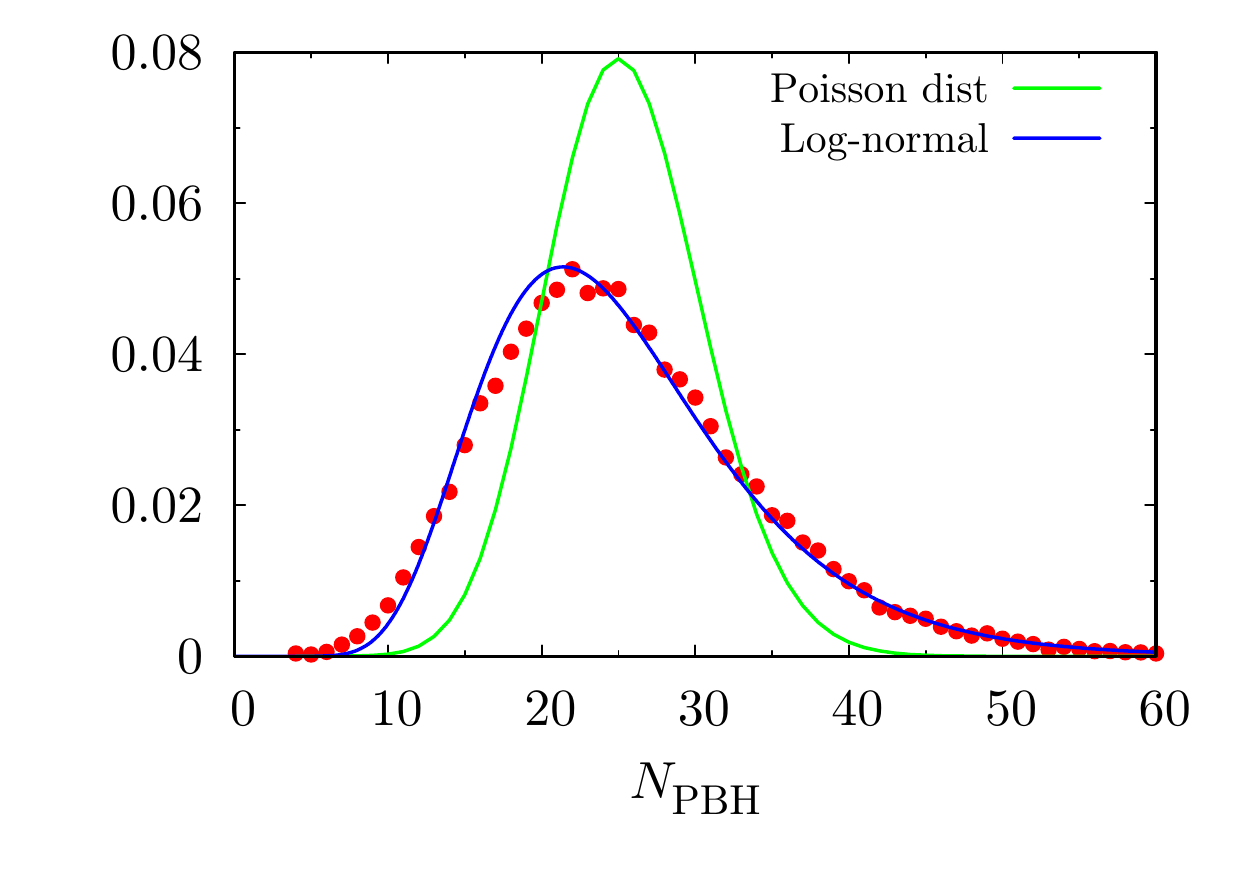}
\label{subfig:dist1a}
}
\subfigure[$n_\delta = 2$]{
\includegraphics [width = 7.5cm, clip]{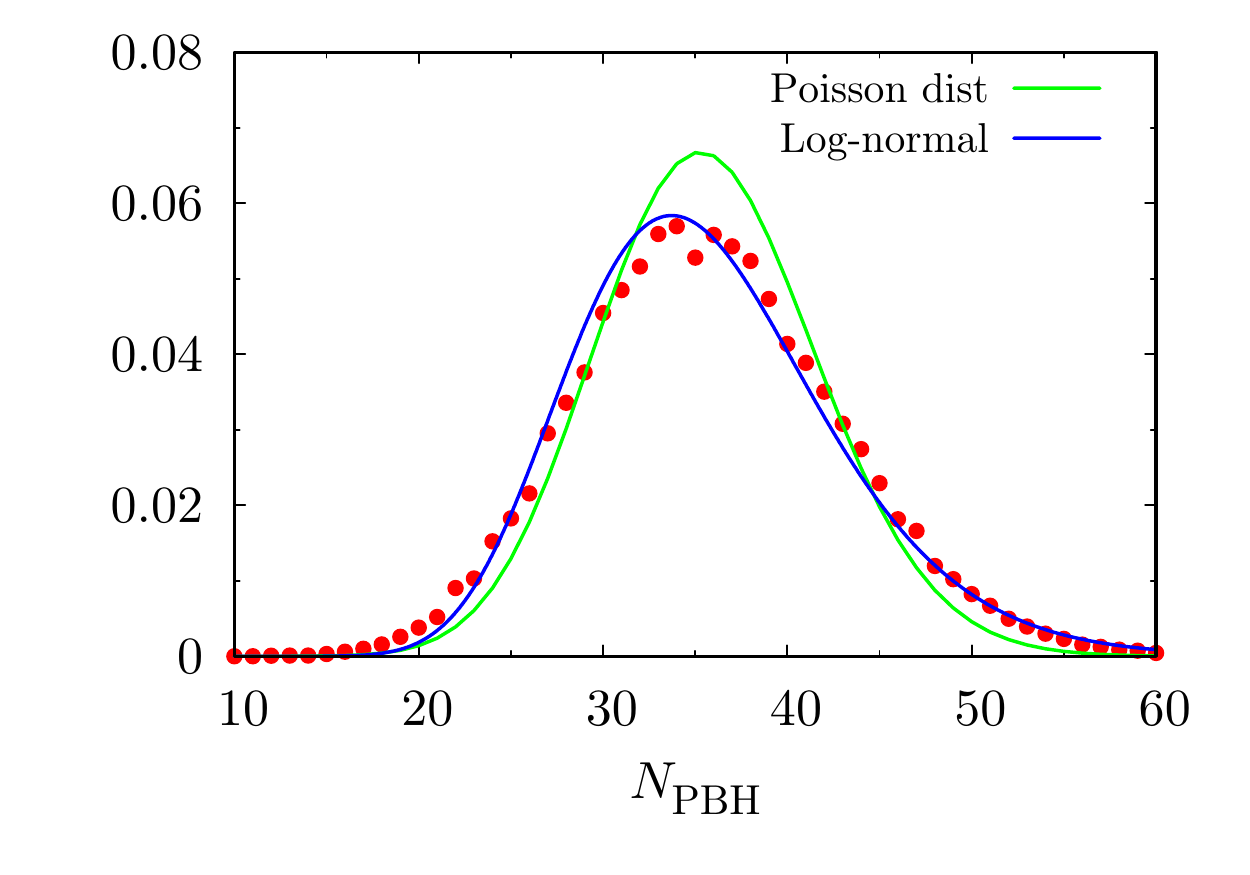}
\label{subfig:dist1b}
}
\subfigure[$n_\delta = 3$]{
\includegraphics [width = 7.5cm, clip]{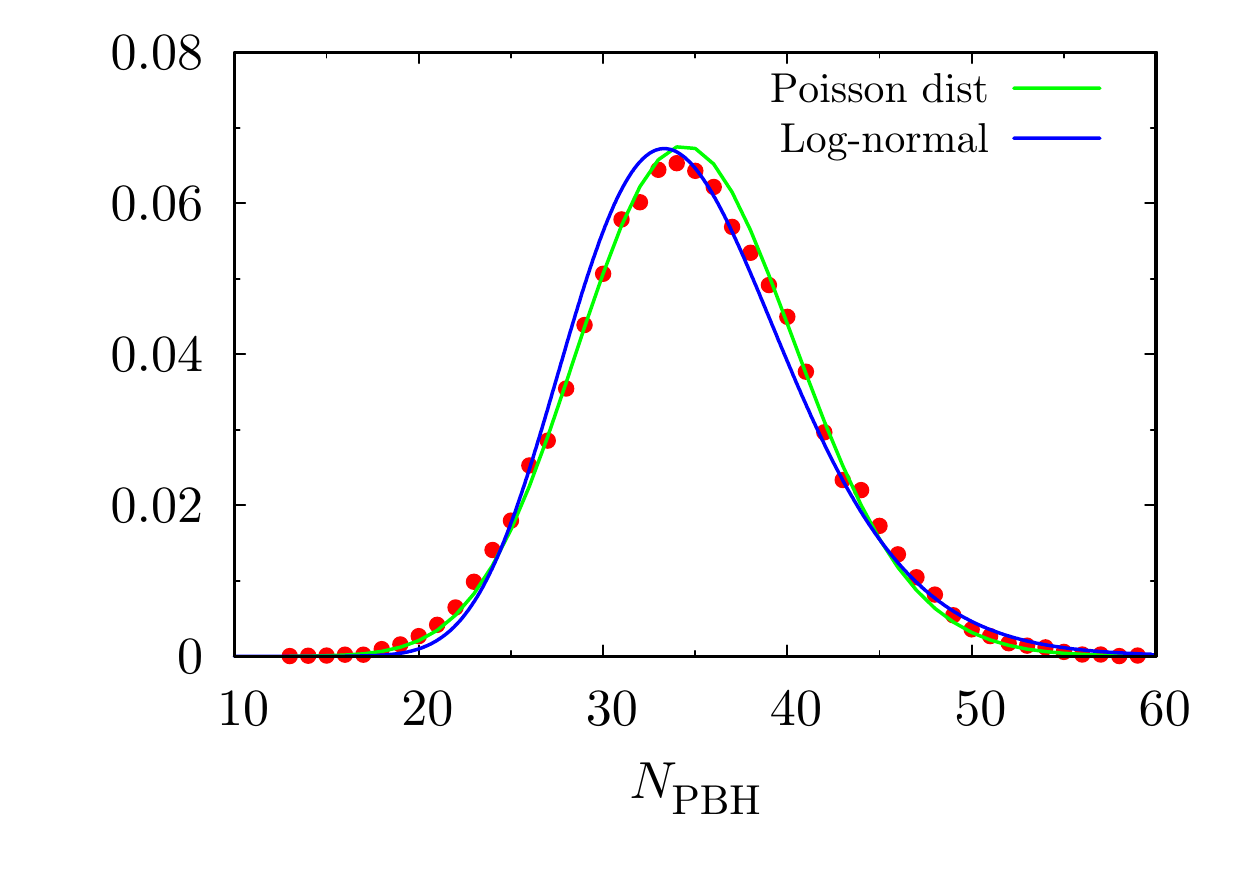}
\label{subfig:dist1c}
}
\subfigure[$n_\delta = 4$]{
\includegraphics [width = 7.5cm, clip]{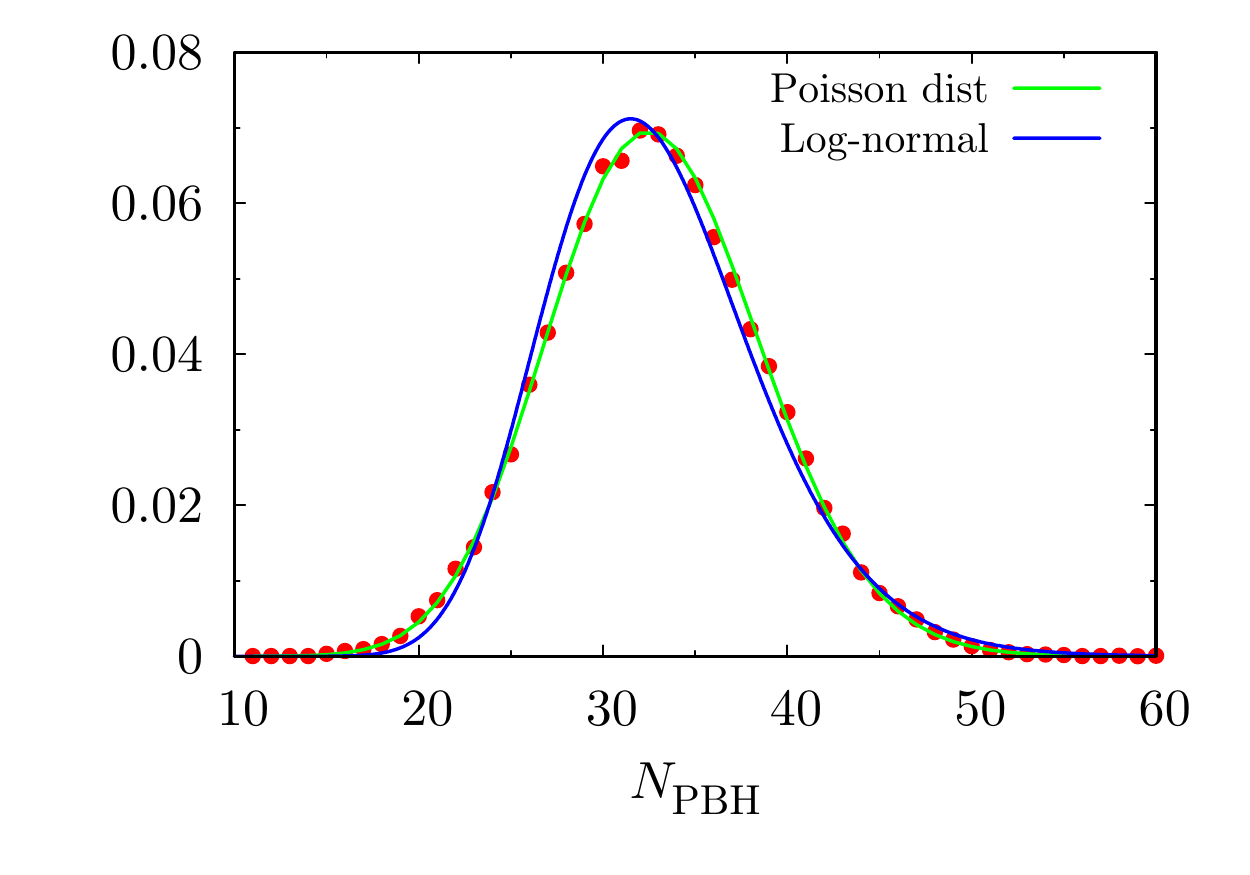}
\label{subfig:dist1d}
}
\caption{
The PBH number distributions for various spectral indices $n_\delta$. The red dots are numerical results and green and blue lines correspond to the Poisson and log-normal distributions respectively. The number of girds in lattice space is $512^3$ and the number of trials is 30,000.
}
\label{fig:dist1}
\end{figure}

\begin{figure}[tp]
\centering
\subfigure[$n_\delta = 1.5$]{
\includegraphics [width = 7.5cm, clip]{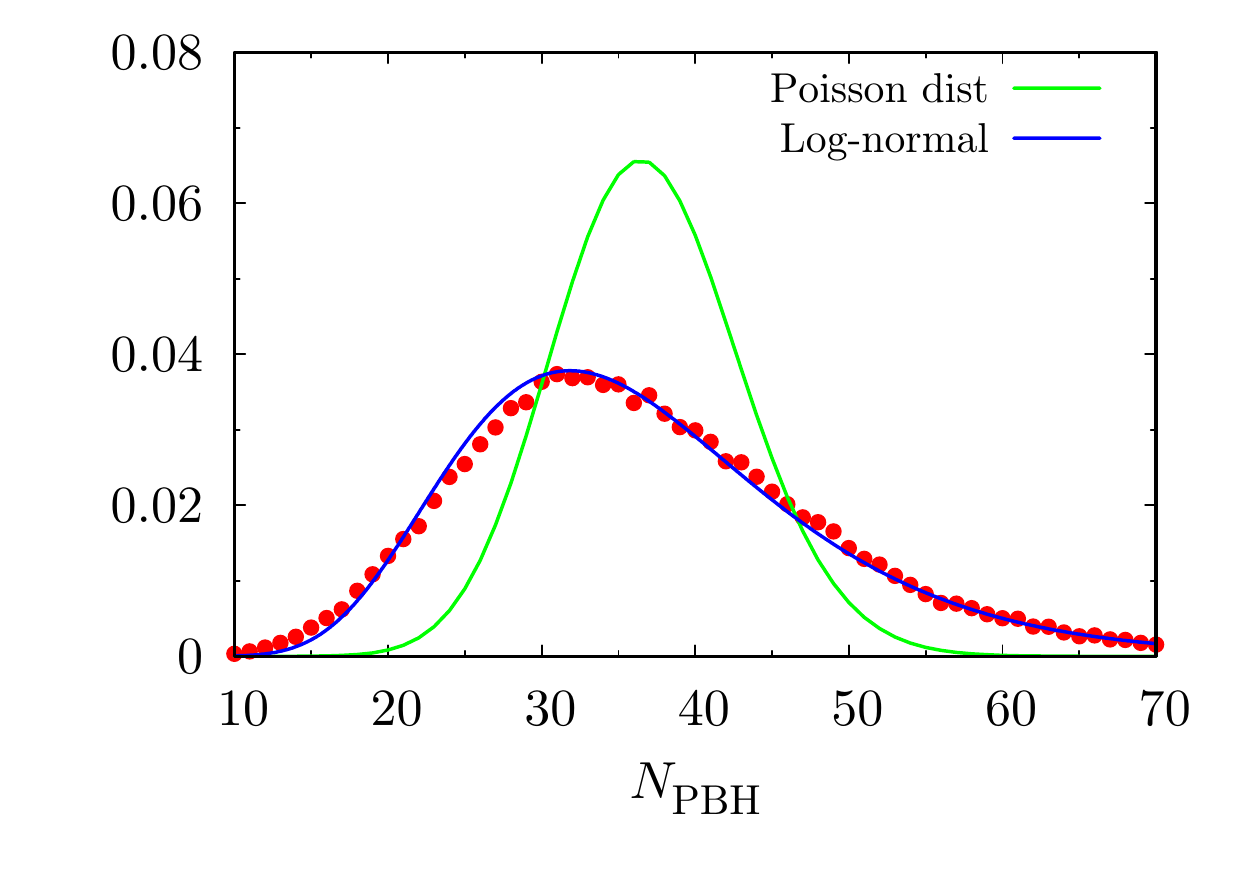}
\label{subfig:dist2a}
}
\subfigure[$n_\delta = 2$]{
\includegraphics [width = 7.5cm, clip]{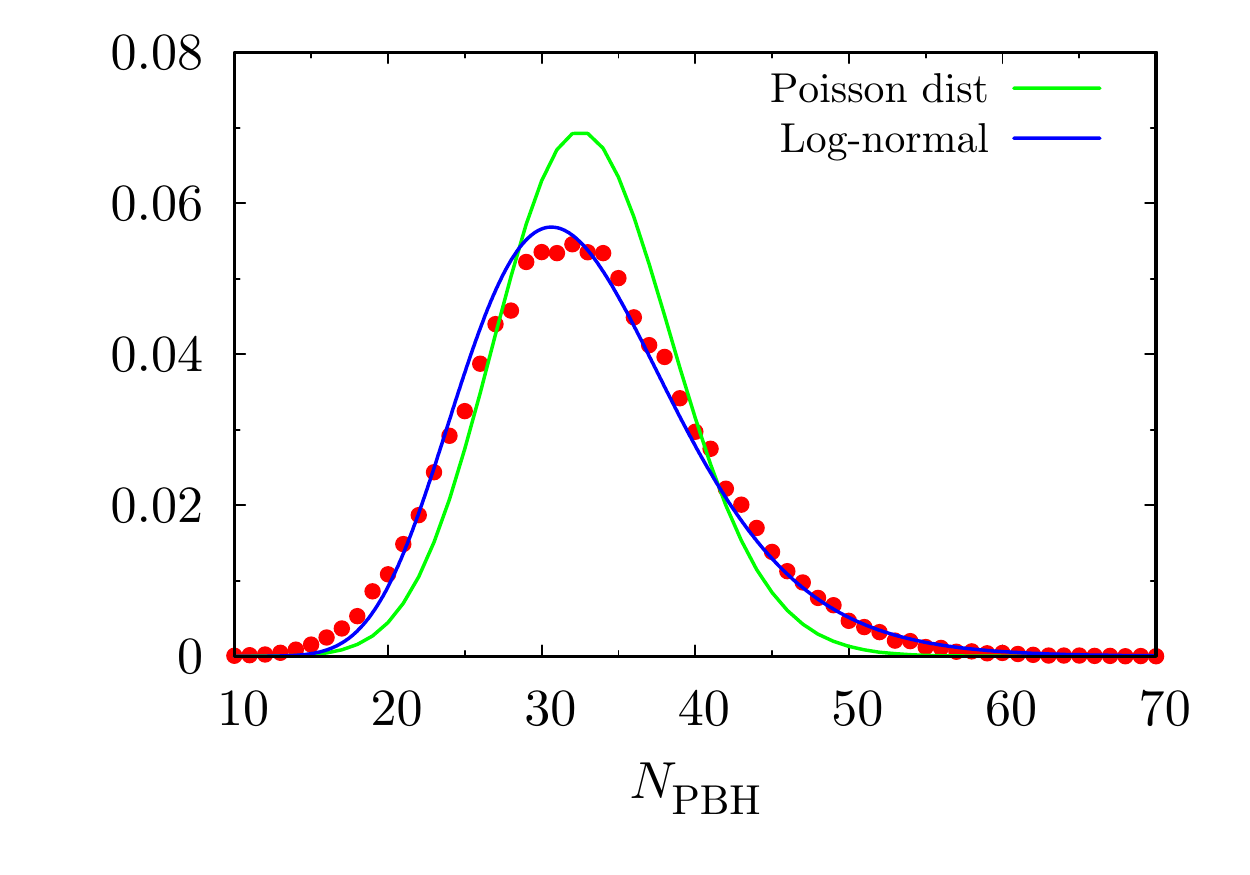}
\label{subfig:dist2b}
}
\subfigure[$n_\delta = 3$]{
\includegraphics [width = 7.5cm, clip]{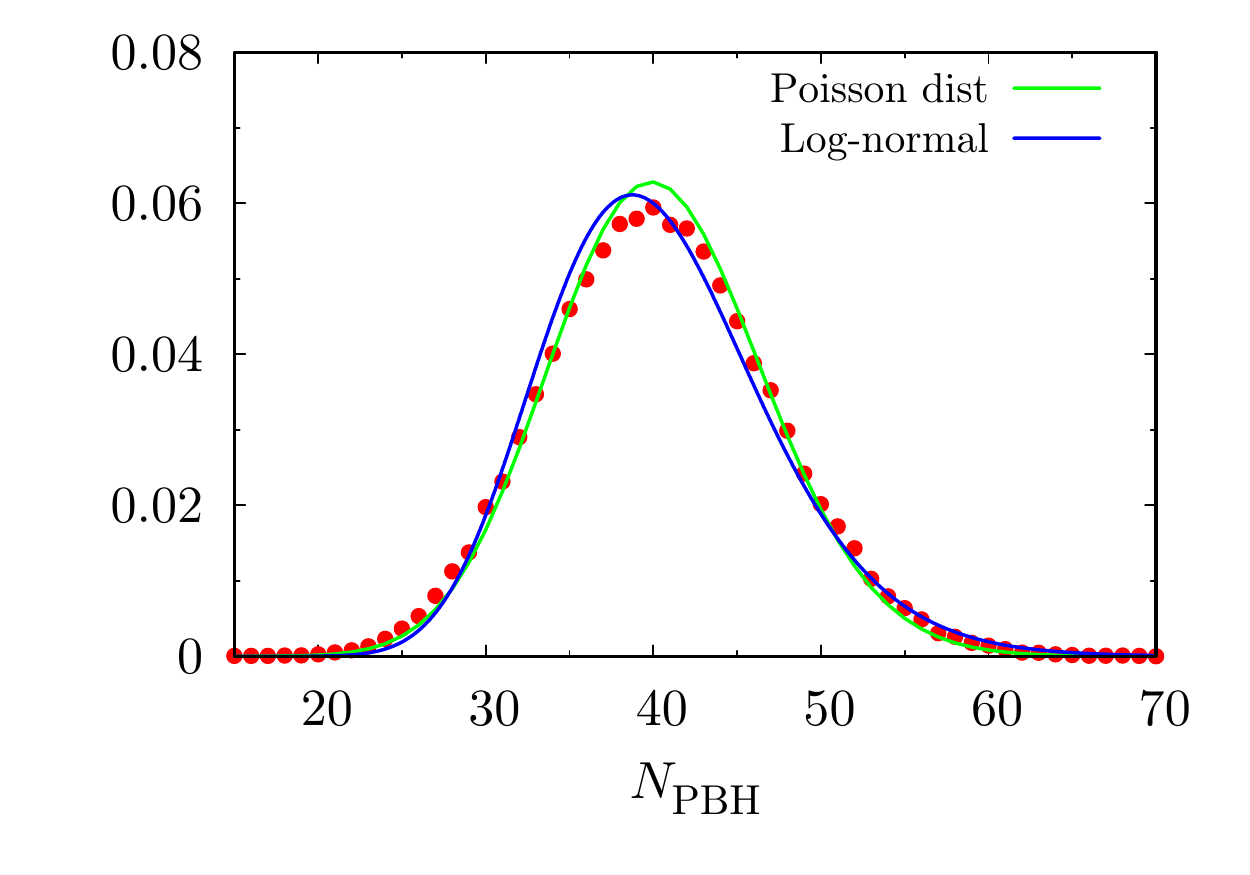}
\label{subfig:dist2c}
}
\subfigure[$n_\delta = 4$]{
\includegraphics [width = 7.5cm, clip]{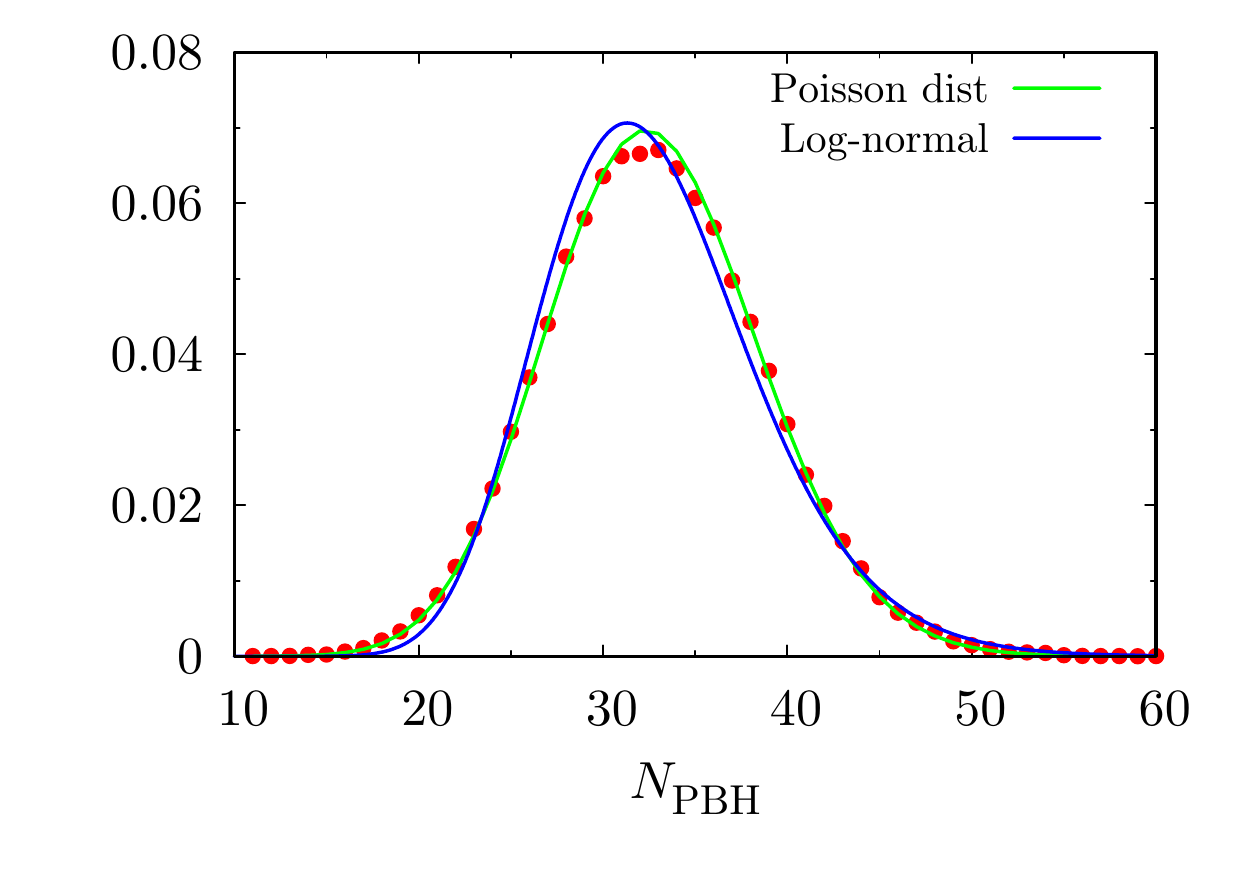}
\label{subfig:dist2d}
}
\caption{
The PBH number distributions for various spectral indices $n_\delta$. The red dots are numerical results and green and blue lines correspond to the Poisson and log-normal distributions respectively. The number of girds in lattice space is $256^3$ and the number of trials is 72,000.
}
\label{fig:dist2}
\end{figure}

We have generated 30,000 (72,000) realizations in the lattice space with $512^3$ ($256^3$) grids and the smoothing scale is set to be $R=L/N_{\rm grid}$.
The resultant distributions of the PBH number are shown in Figure~\ref{fig:dist1} and \ref{fig:dist2}, where in each panel we have taken different value of the spectral index $n_\delta$. We have found, for $n_\delta=1.5$ and $2$, the distribution shows a good agreement with the log-normal distribution while it follows the Poisson distribution for $n_\delta=3$ and 4.

\begin{figure}[tp]
\centering
\includegraphics [width = 9cm, clip]{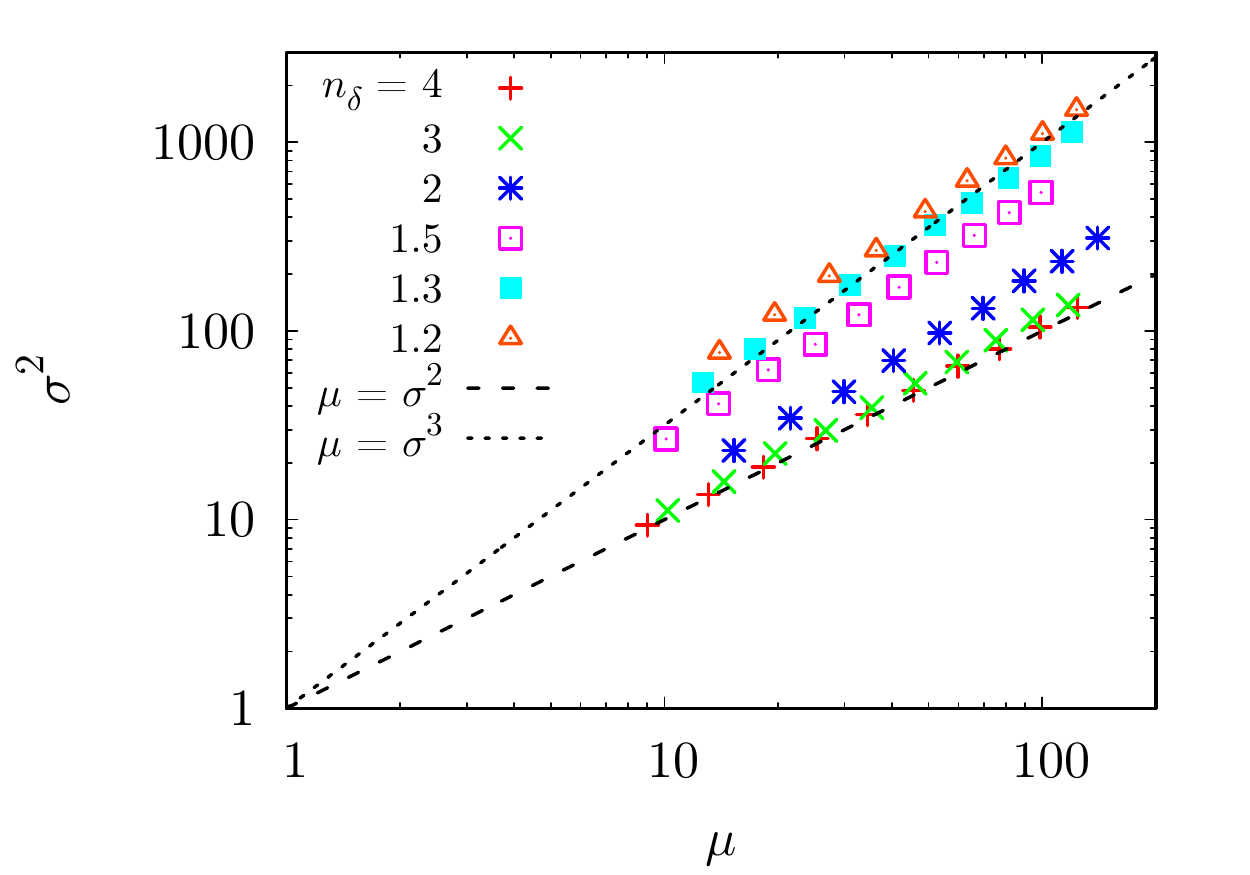}
\caption{
Plot in $\mu$-$\sigma^2$ plane for various $n_\delta$. The black dashed and dotted lines correspond to $\sigma^2 = \mu^{3/2}$ (log-normal) and $\sigma^2 = \mu$ (Poisson) respectively.
}
\label{fig:mu-var}
\end{figure}

Figure~\ref{fig:mu-var} shows the relation between the mean value and the variance for various values of $n_\delta$. It shows that there is a monotonic relation : $\sigma^2 = \alpha \mu^p$. In the case with the Poisson distribution, the relation is $\sigma^2 = \mu$. and the cases with $n_\delta = 3$ and 4 show this relation.
On the other hand, in the case with $n_\delta=1.2$\,--\,2, $p$ significantly deviates from 1 and it approaches 3/2 as $n_\delta \to 1$. This behavior can be roughly understood as shown in the following subsection.

\subsubsection{Log-normal distribution of the PBH number}

\begin{figure}[tp]
\centering
\includegraphics [width = 14cm, clip]{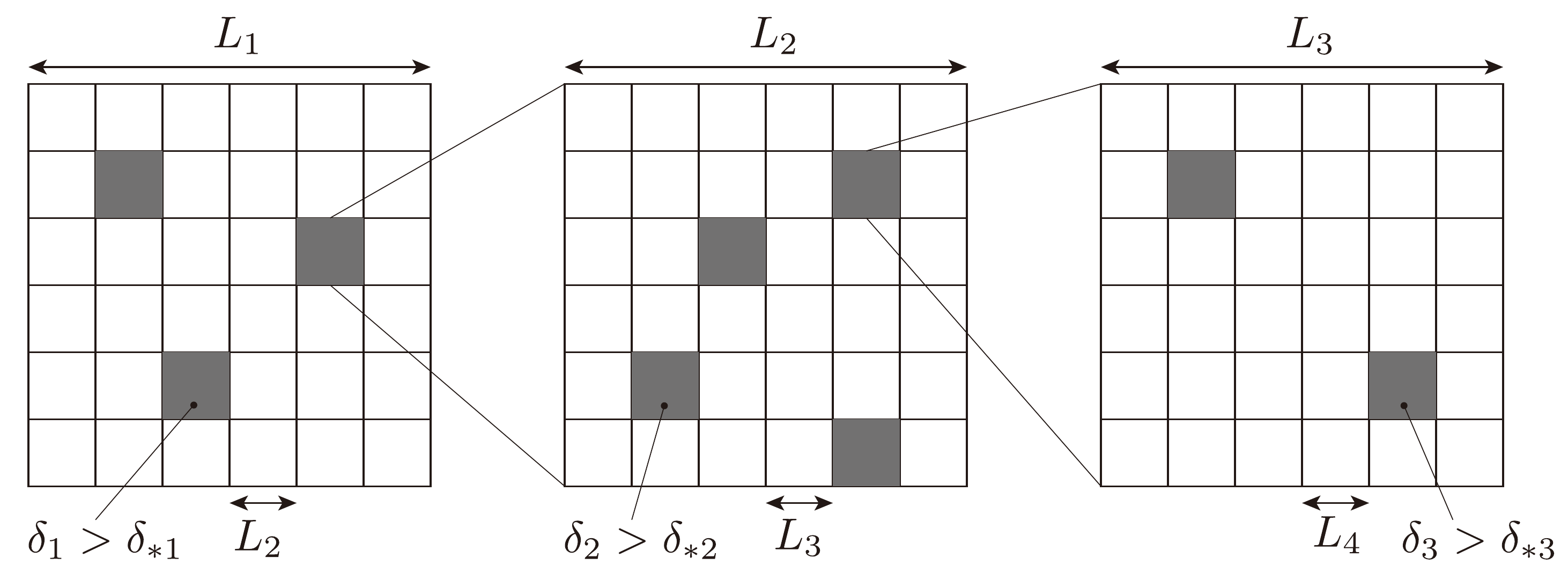}
\caption{
An illustration of counting the region where the PBH formation occurs.
}
\label{fig:PBHcount}
\end{figure}

Here we give an intuitive understanding for the log-normal distribution of the number of PBHs and the scaling relation : $\sigma^2 \propto \mu^{3/2}$.
As one can see in Fig.\,\ref{fig:deltas2}, the PBH formation occurs, or $\delta_s > 0.4$, where larger-scale fluctuation is relatively large especially for smaller $n_\delta$.
To understand the effect of the larger scale fluctuations, let us first consider a subspace of the Universe with length scale $L_1$ which is much larger than the horizon scale at the PBH formation. Then, we divide the subspace into smaller subspaces with length scale $L_2$ and count the region where the smoothed density contrast over the scale, $\delta_1$, has a local peak which is larger than a certain threshold value, $\delta_{*1}$. We repeat the same procedure for each subspace satisfying $\delta_i > \delta_{*i}$ until the length scale becomes comparable to the scale of the PBH formation as illustrated in Figure~\ref{fig:PBHcount}. Note that one can choose the value of each $\delta_{*i}$ and $L_i$ so that long-wavelength density profile can enhance smaller-scale local overdensities. We choose the threshold value in such a way that $\delta_{*1} < \delta_{*2} < \dots < \delta_{*n} = \delta_{\rm cr}$, and then the resultant PBH number is obtained as 
\begin{equation}
N_{\rm PBH} = N_{\delta_1 > \delta_{*1}} \times N_{\delta_2 > \delta_{*2}} \times \dots \times N_{\delta_n > \delta_{*n}} \,.
\end{equation}

It is known that, from the central limit theorem, the distribution of the product of a number of random variables, regardless of the distribution function of each individual variable, follows the log-normal distribution. Hence, one obtains the distribution function for the number of PBHs as follows:
\begin{equation}
\mathbb{P}(N_{\rm PBH}) = \frac{1}{\sqrt{2 \pi} \sigma_{\log} N_{\rm PBH}} 
\exp \left[\frac{(\log N_{\rm PBH}-\mu_{\log})^2}{2 \sigma_{\log}^2} \right] \, ,
\end{equation}
where $\mu_{\log}$ and $\sigma_{\log}$ are related to the mean value $\mu$ and the variance $\sigma^2$ via
\begin{align}
\mu & \equiv \left\langle N_{\rm PBH} \right\rangle = \bar{N}_{\rm PBH}
= \exp \left( \mu_{\log}+\frac{1}{2}\sigma^2_{\log} \right) \, , 
\\[1mm]
\sigma^2 & \equiv \left\langle \left( N_{\rm PBH} - \bar{N}_{\rm PBH} \right)^2 \right\rangle
= \exp \left( 2\mu_{\log}+\sigma^2_{\log} \right) \left[ \exp \left( \sigma^2_{\log} \right) - 1 \right] \, ,
\end{align}
where angular brackets and overbar denote the ensemble average.

Here we choose $\delta_{*i}$ for each $i$ such that the probability to find the region with $\delta_i > \delta_{*i}$ is small enough and $N_i \equiv N_{\delta_i > \delta_{*i}} \sim \mathcal{O}(1)$. In this case, each $N_i$ obeys the Poisson distribution:
\begin{equation}
\mathbb{P}(N_i) = \frac{\lambda_i^{N_i} e^{-\lambda_i}}{N_i!} \, ,
\end{equation}
where $\lambda_i = \sum_{N_i=0}^{\infty} N_i \mathbb{P}(N_i) = \sum_{N_i=0}^{\infty}(N_i-\lambda_i)^2 \mathbb{P}(N_i)$ is the expectation value and also the variance.
One can calculate the expectation value of $N_{\rm PBH} = \prod_{i=1}^{n} N_i$ by using a joint distribution function:
\begin{equation}
\mathbb{P}(N_1,N_2,\cdots,N_n) = \prod_{i=1}^{n} \mathbb{P}(N_i) \, . 
\end{equation}
The expectation value can be simply obtained as
\begin{equation}
\mu =\sum_{N_1,\dots, N_n} N_1 \dots N_n \mathbb{P}(N_1, \cdots , N_n) = \prod_{i=1}^{n} \sum_{N_i=0}^{\infty} N_i \mathbb{P}(N_i) = \prod_{i=1}^{n} \lambda_i \, ,
\end{equation}
and the variance can be obtained as
\begin{equation}
\sigma^2 = \sum_{N_1,\dots, N_n} \left( \prod_{i} N_i - \prod_{i} \lambda_i \right)^2 \mathbb{P}(N_1, \cdots , N_n) =\prod_{i} \left( \lambda_i^2 + \lambda_i \right) - \prod_{i} \lambda_i^2 \, ,
\end{equation}
where we have used $\sum_{N=0}^{\infty} N^2 \mathbb{P}(N,\lambda) = \lambda^2 + \lambda$ for the Poisson distribution.

Let us further simplify the situation by properly choosing $\delta_{*i}$ such that $\lambda_1 = \lambda_2 = \dots = \lambda_n \equiv \lambda$. In this case, we have 
\begin{equation}
\sigma^2 = \lambda^n (\lambda+1)^n - \lambda^{2n} \, .
\end{equation}
Assuming $\lambda \sim \mathcal{O}(1)$ and $n \gg1$, the largest contribution in $(\lambda+1)^n$ is $n!/[(n/2)!]^2 \lambda^{n/2}$. This is justified by the fact that PBH formation is very rare and $\lambda$ cannot be significantly larger than $\mathcal{O}(1)$. Then, one obtains the scaling relation as 
\begin{equation} 
\label{eq:sigma2-mu}
\sigma^2 \propto \lambda^{3n/2} \propto \mu^{3/2} \, .
\end{equation}

\subsection{Matter power spectrum and halo mass function}

\begin{figure}[tp]
\centering
\subfigure[$M_{\rm PBH} = M_\odot$]{
\includegraphics [width = 7.5cm, clip]{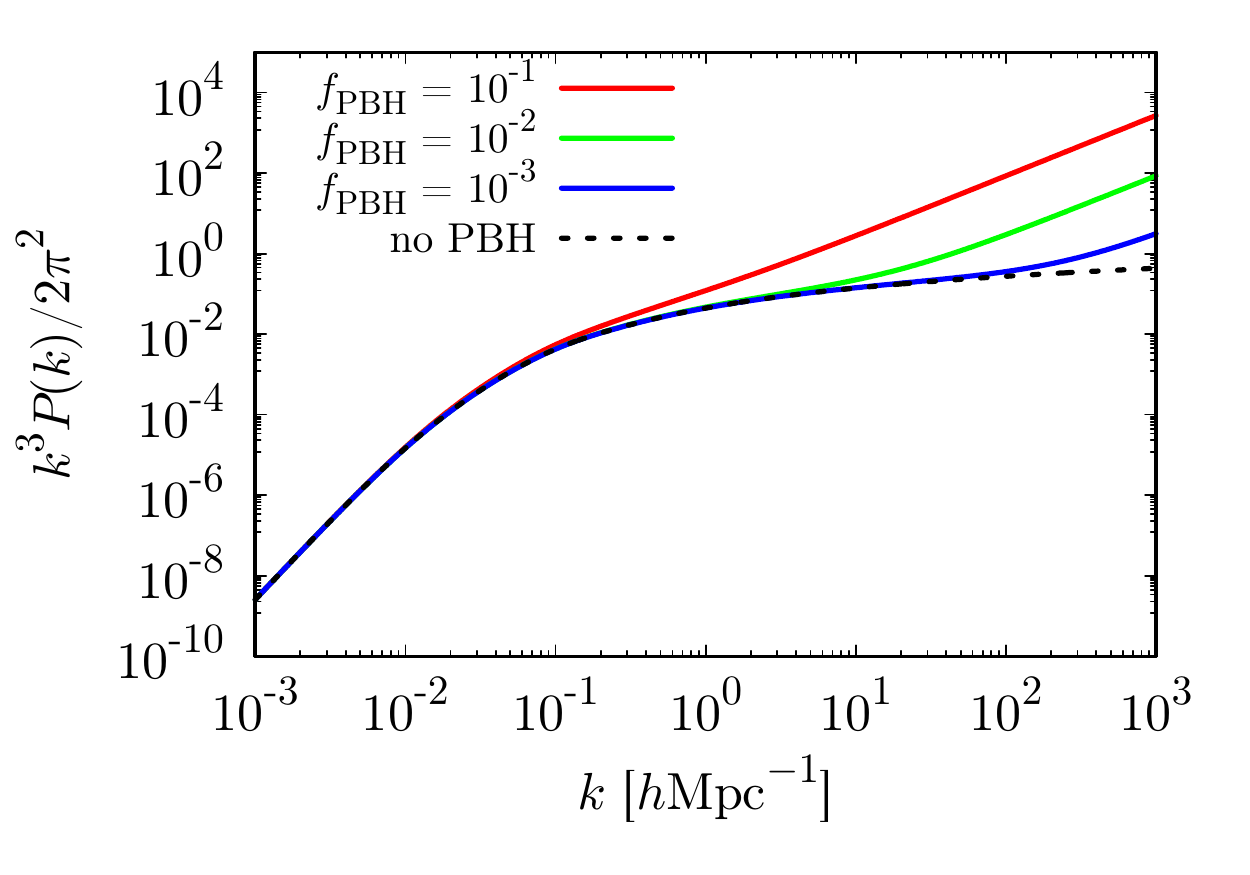}
\label{subfig:ps1}
}
\subfigure[$M_{\rm PBH} = 10M_\odot$]{
\includegraphics [width = 7.5cm, clip]{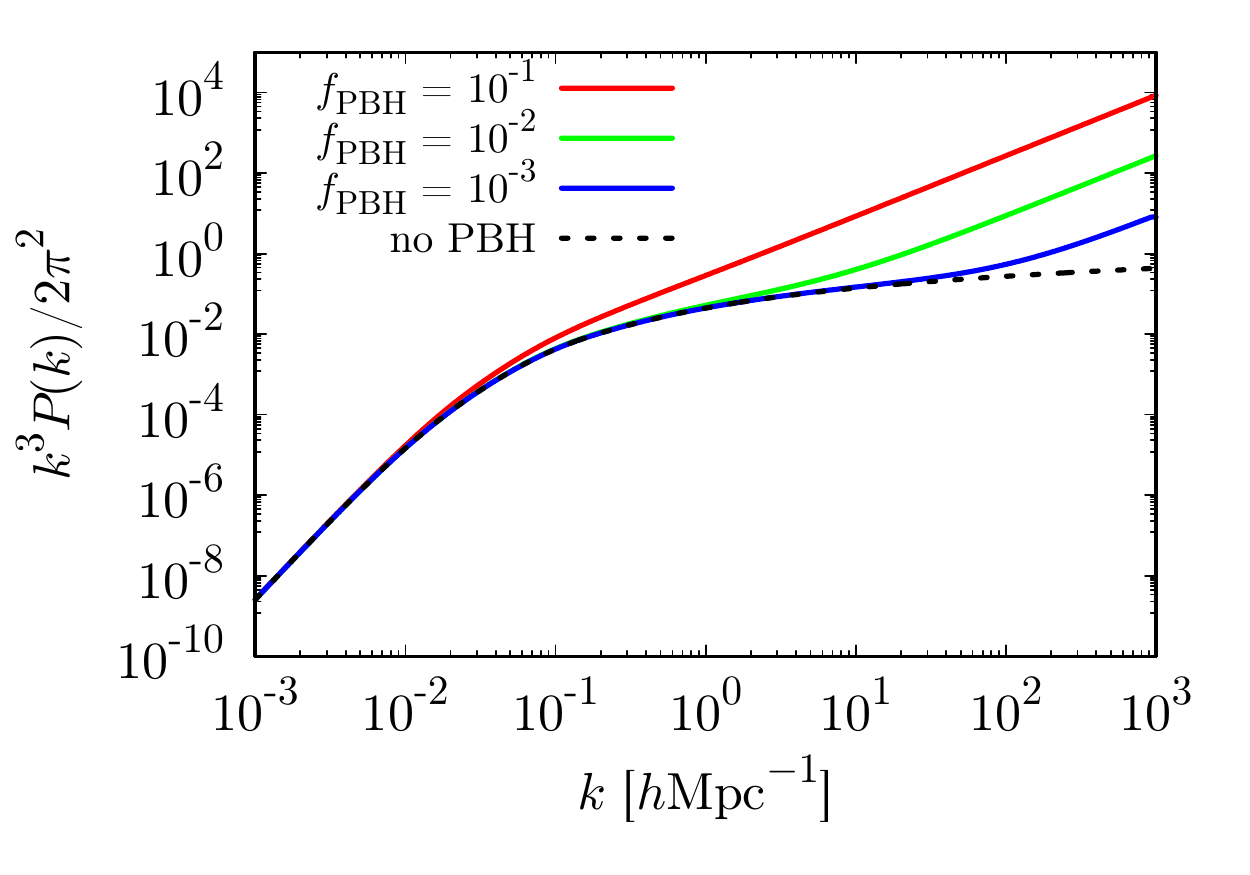}
\label{subfig:ps2}
}
\subfigure[$M_{\rm PBH} = 100M_\odot$]{
\includegraphics [width = 7.5cm, clip]{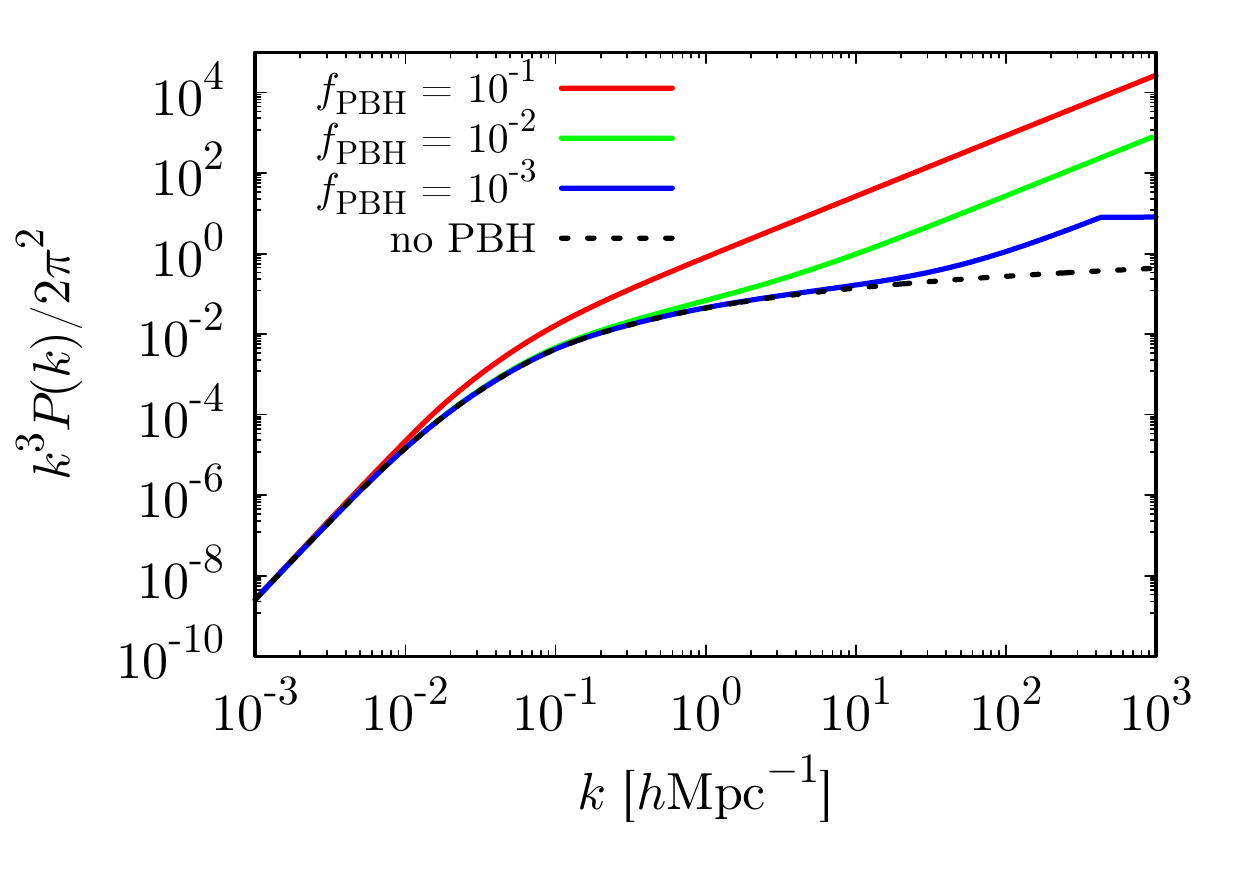}
\label{subfig:ps3}
}
\subfigure[$M_{\rm PBH} = 1000M_\odot$]{
\includegraphics [width = 7.5cm, clip]{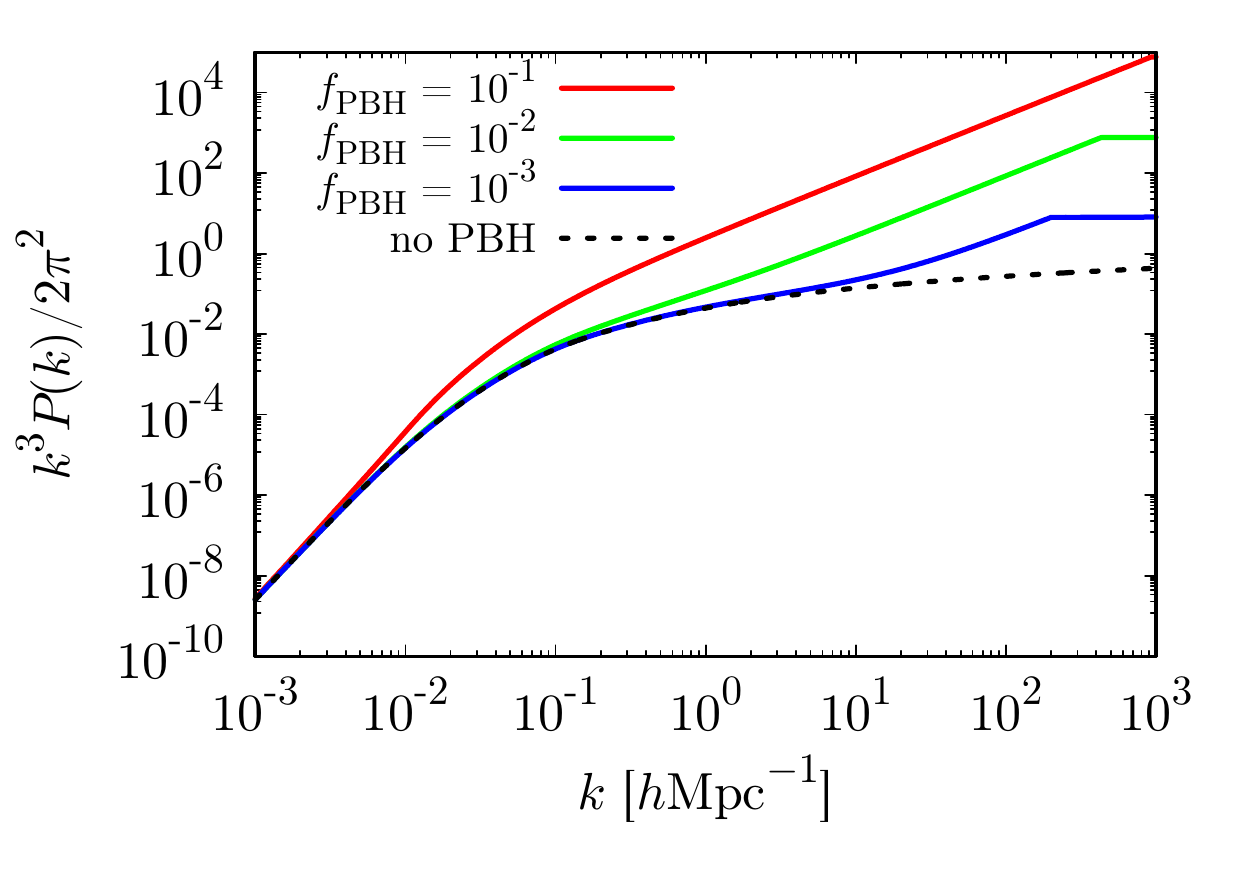}
\label{subfig:ps4}
}
\caption{
The linear matter power spectrum. We have taken $\alpha=1$, $z=10$, $M_{\rm PBH} = M_\odot$ (a), $10M_\odot$ (b), $100M_\odot$ (c), $1000M_\odot$ (d) and $f_{\rm PBH} = 10^{-1}$, $10^{-2}$ and $10^{-3}$ from top to bottom. The dashed line shows the linear power spectrum without PBH isocurvature fluctuations.
}
\label{fig:powerspec}
\end{figure}

The fluctuation of the PBH number gives an additional contribution to the late-time matter density contrast. Focusing only on the log-normal fluctuation, the variance of the density contrast of PBHs is  
\begin{equation}
\label{eq:delta2PBH}
\langle \delta_{\rm PBH}^2 \rangle = \frac{\left\langle \left( N_{\rm PBH}-\bar{N}_{\rm PBH} \right)^2 \right\rangle}{\bar{N}_{\rm PBH}^2} = 
\frac{\alpha \bar{N}_{\rm PBH}^{3/2}}{\bar{N}_{\rm PBH}^2}
=\frac{\alpha}{\bar{N}_{\rm PBH}^{1/2}} \, ,
\end{equation}
where we have used the scaling law obtained in the previous subsection: 
$\sigma^2 = \alpha \bar{N}_{\rm PBH}^p$ with $p=3/2$.
The number of PBHs in a given comoving volume $V$ is
\begin{equation}
\label{eq:NPBH}
\bar{N}_{\rm PBH} = \frac{\bar\rho_{\rm PBH}(z) V}{M_{\rm PBH}(1+z)^3} = 3.3 \times 10^{10} f_{\rm PBH} \left( \frac{M_\odot}{M_{\rm PBH}} \right) \left( \frac{V}{{\rm Mpc}^3} \right) \, ,
\end{equation}
where 
$\bar{\rho}_{\rm PBH}(z) = f_{\rm PBH} \Omega_{\rm CDM} \rho_{{\rm cr},0}(1+z)^3$ 
is the homogeneous energy density of PBHs with 
$\rho_{{\rm cr},0} = 2.78 \times 10^{11}M_\odot \, h^2 \, {\rm Mpc}^{-3}$ being the critical energy density today.
Using the relation $V = (2\pi/k)^3$ with $k$ being a comoving wave number and taking into account the linear growth factor $D(z)$ and the transfer function for the isocurvature perturbation $T_{\rm iso}(k)$, with the initial PBH spectrum $P_{\rm PBH} = V \left\langle \delta_{\rm PBH}^2 \right\rangle$, the isocurvature contribution to the matter power spectrum at redshift $z$ is given by
\begin{equation}
\begin{split}
P_{\rm iso}(k,z) &= f_{\rm PBH}^2 
T^2_{\rm iso}(k) D^2(z) P_{\rm PBH}
\\[1mm]
&= 8.6 \times 10^{-5} ~{\rm Mpc}^3~\left( \frac{k}{{\rm Mpc}^{-1}} \right)^{-3/2} \alpha f_{\rm PBH}^{3/2} \left( \frac{M_{\rm PBH}}{M_\odot} \right)^{1/2} T^2_{\rm iso}(k) D^2(z) \, .
\end{split}
\end{equation}
Note that there is an upper limit on the power spectrum. For $\bar{N}_{\rm PBH} < 1$, the fluctuation $\sim \alpha^{1/2}\bar{N}_{\rm PBH}^{3/4}$ becomes larger than the mean value, $\bar{N}_{\rm PBH}$, and \eqref{eq:delta2PBH} is no longer valid in that case but rather $\langle \delta^2_{\rm PBH} \rangle \sim 1$.
Then, let us define the wavenumber $k_{\rm nl}$ so as to satisfy $\langle \delta^2_{\rm PBH} \rangle = 1$. It is given by
\begin{equation}
k_{\rm nl} = 2.0 \times 10^4 \alpha^{-2/3} f_{\rm PBH}^{1/3} \left(\frac{M_\odot}{M_{\rm PBH}} \right)^{1/3} \, ,
\end{equation}
and the power spectrum is given by\footnote{
On scales $k > k_{\rm nl}$, the density contrast of PBH becomes non-linear, which can yield the non-zero bispectrum of the isocurvature perturbation. It can predict sizable value of isocurvature-type $f_{\rm NL}$ on small scales. 
}
\begin{equation}
P_{\rm iso}(k,z) \to
\begin{cases} 
P_{\rm iso}(k,z) & \text{for} \quad k<k_{\rm nl} \, ,
\\[1mm]
P_{\rm iso}(k_{\rm nl},z) & \text{otherwise} \, .
\end{cases}
\end{equation}
Figure~\ref{fig:powerspec} shows the linear matter power spectrum as the sum the adiabatic and isocurvature contributions. The isocurvature component can be dominant on small scales even if PBHs account for only a small fraction of total dark matter.

\begin{figure}[tp]
\centering
\subfigure[$M_{\rm PBH} = M_\odot$]{
\includegraphics [width = 7.5cm, clip]{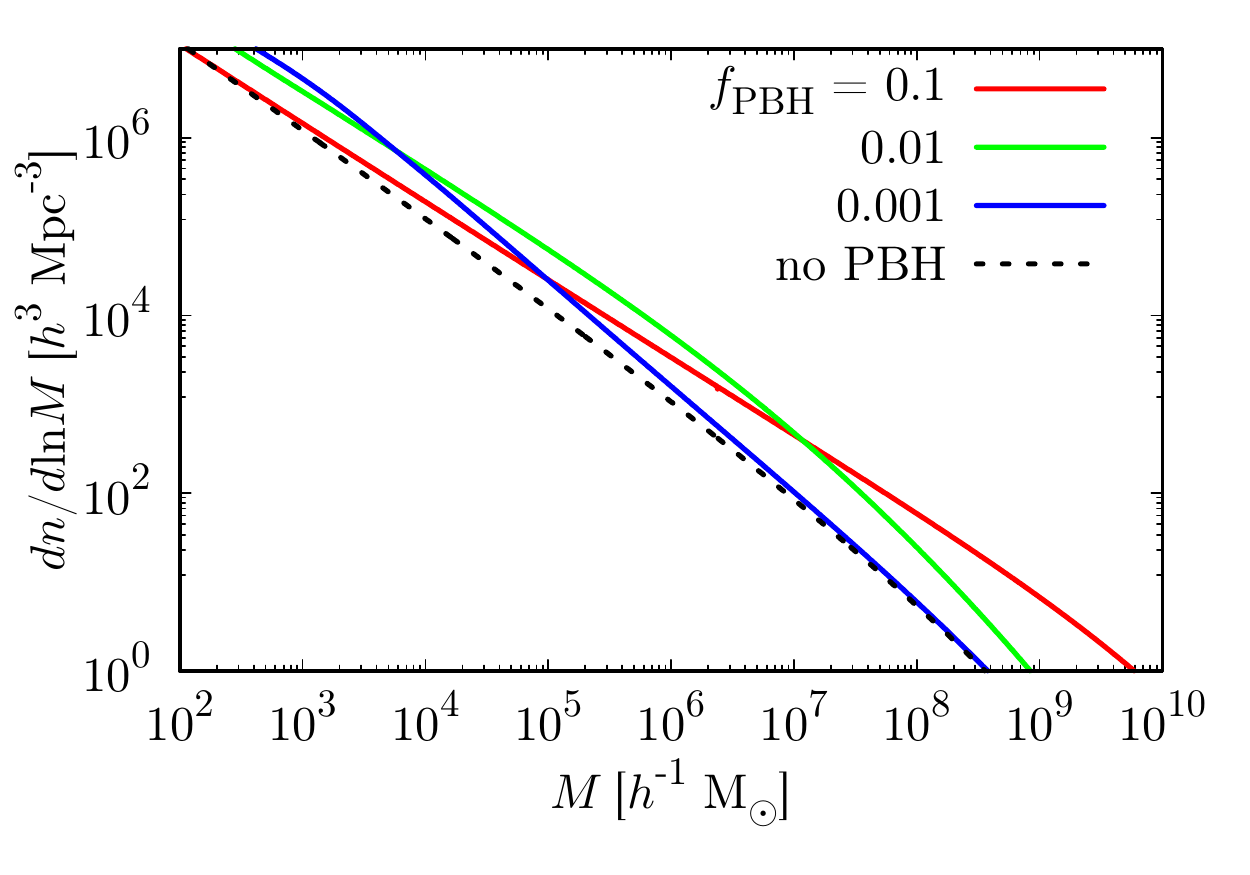}
\label{subfig:mf1}
}
\subfigure[$M_{\rm PBH} = 10M_\odot$]{
\includegraphics [width = 7.5cm, clip]{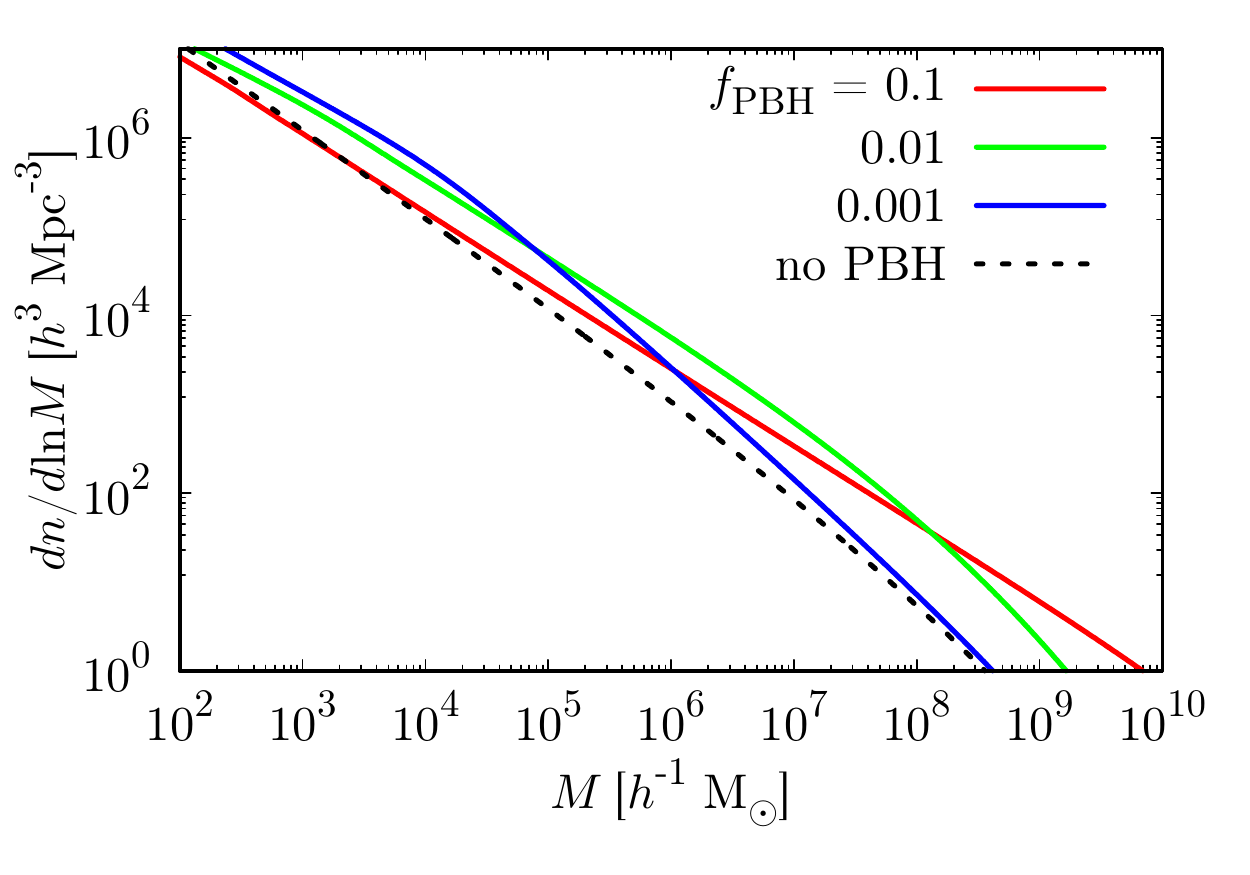}
\label{subfig:mf2}
}
\subfigure[$M_{\rm PBH} = 100M_\odot$]{
\includegraphics [width = 7.5cm, clip]{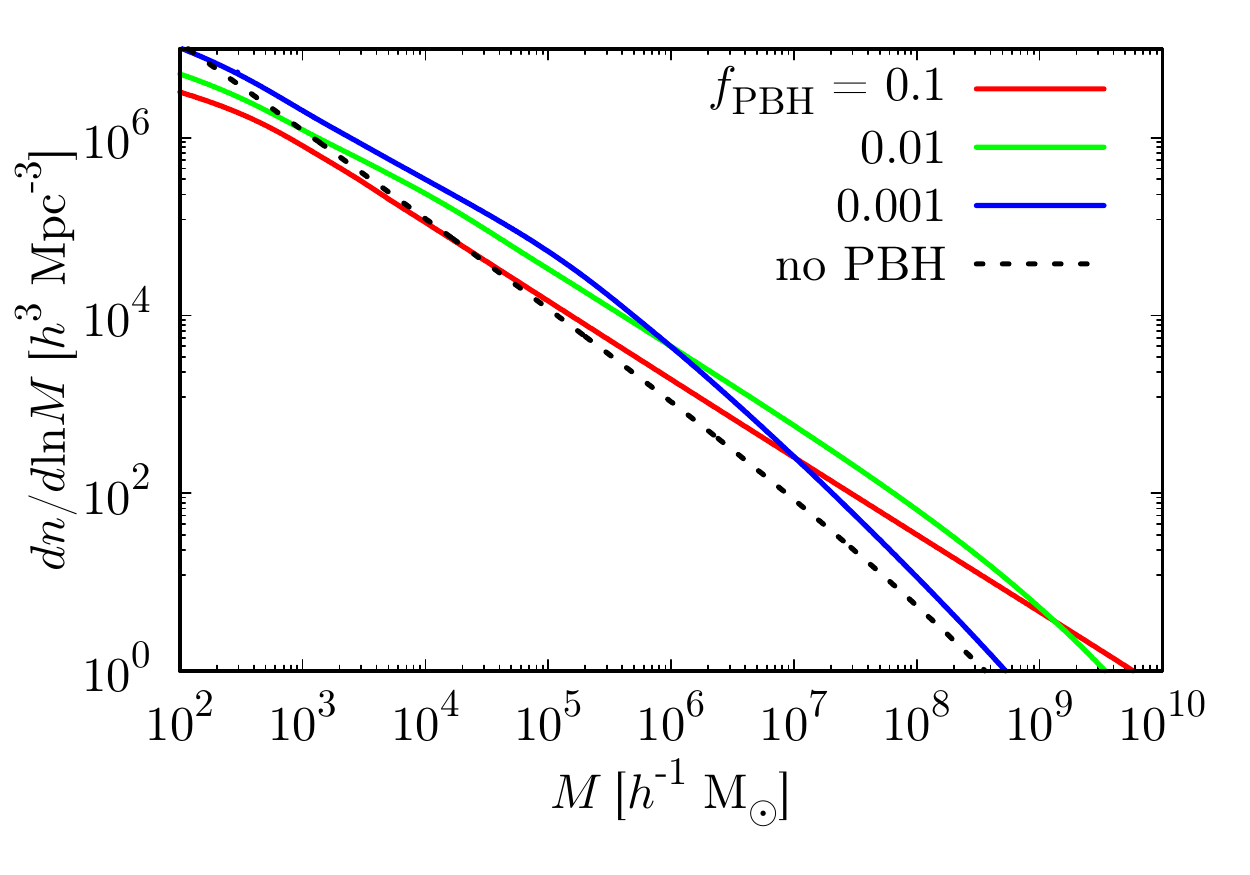}
\label{subfig:mf3}
}
\subfigure[$M_{\rm PBH} = 1000M_\odot$]{
\includegraphics [width = 7.5cm, clip]{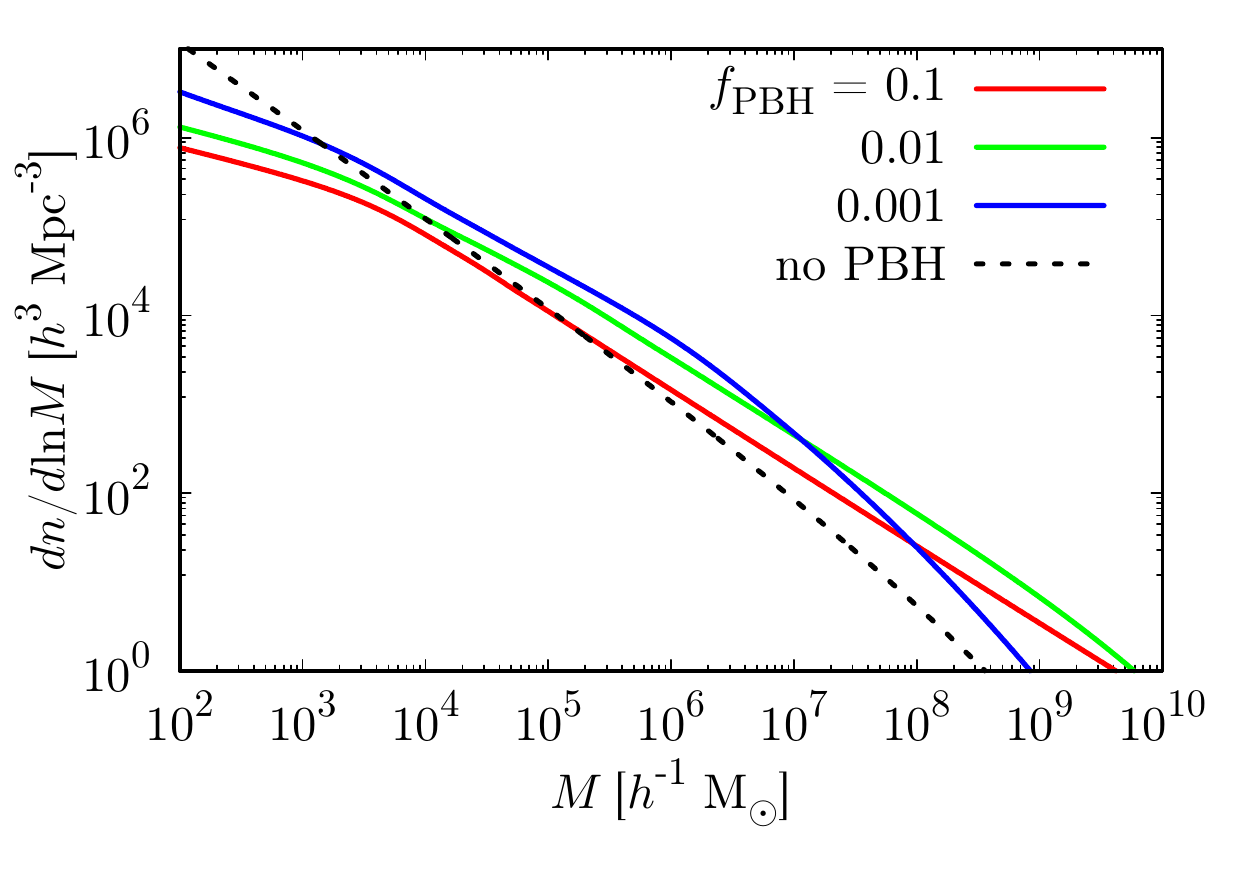}
\label{subfig:mf4}
}
\caption{
The halo mass function. We have taken $\alpha=1$, $z=10$, $M_{\rm PBH} = M_\odot$ (a), $10M_\odot$ (b), $M_{\rm PBH} = 100M_\odot$ (c), $1000M_\odot$ (d) and $f_{\rm PBH} = 10^{-1}$, $10^{-2}$ and $10^{-3}$ from top to bottom. The dashed line shows the linear power spectrum without PBH isocurvature fluctuations.
}
\label{fig:mf}
\end{figure}

Because the matter power spectrum is modified significantly on small scales, the halo mass function can also be modified accordingly. The halo mass function, $dn/dM$, is calculated as
\begin{equation}
\frac{dn}{dM} = \frac{\rho_m}{M} \frac{d\log \sigma^{-1}}{dM} f(\sigma) \, ,
\end{equation}
where $\rho_m$ is the energy density of the matter component and $\sigma$ is given by the formula \,(\ref{eq:sigma2R}) with $M=4\pi \rho_mR^3/3$.
$f(\sigma)$ is an universal fitting function and here we adopt the formula derived in \cite{Sheth:1999mn}. Note that $P_\delta(k)$ in \eqref{eq:sigma2R} is now given by the matter power spectrum which is the sum of the adiabatic and isocurvature contributions. 
Figure~\ref{fig:mf} shows the mass function in the presence of the log-normal fluctuations of the PBH number. 
It significantly deviates from that without PBHs. In addition, it shows a non-trivial dependence on $f_{\rm PBH}$. For example, for $M_{\rm PBH} = M_\odot$ and $10M_\odot$, the number of smaller haloes decreases as $f_{\rm PBH}$ increases, while the opposite happens for the number of larger haloes. It comes form the fact that, given a fixed total mass, the increase of larger haloes leads to the decrease of smaller haloes, or smaller haloes are eaten by larger haloes \cite{Sekiguchi:2013lma}.

\section{Constraints on the PBH abundance} 
\label{sec:constraint}

\subsection{Current constraint by the Ly-$\alpha$ forest}

\begin{figure}[tp]
\centering
\includegraphics [width = 12cm, clip]{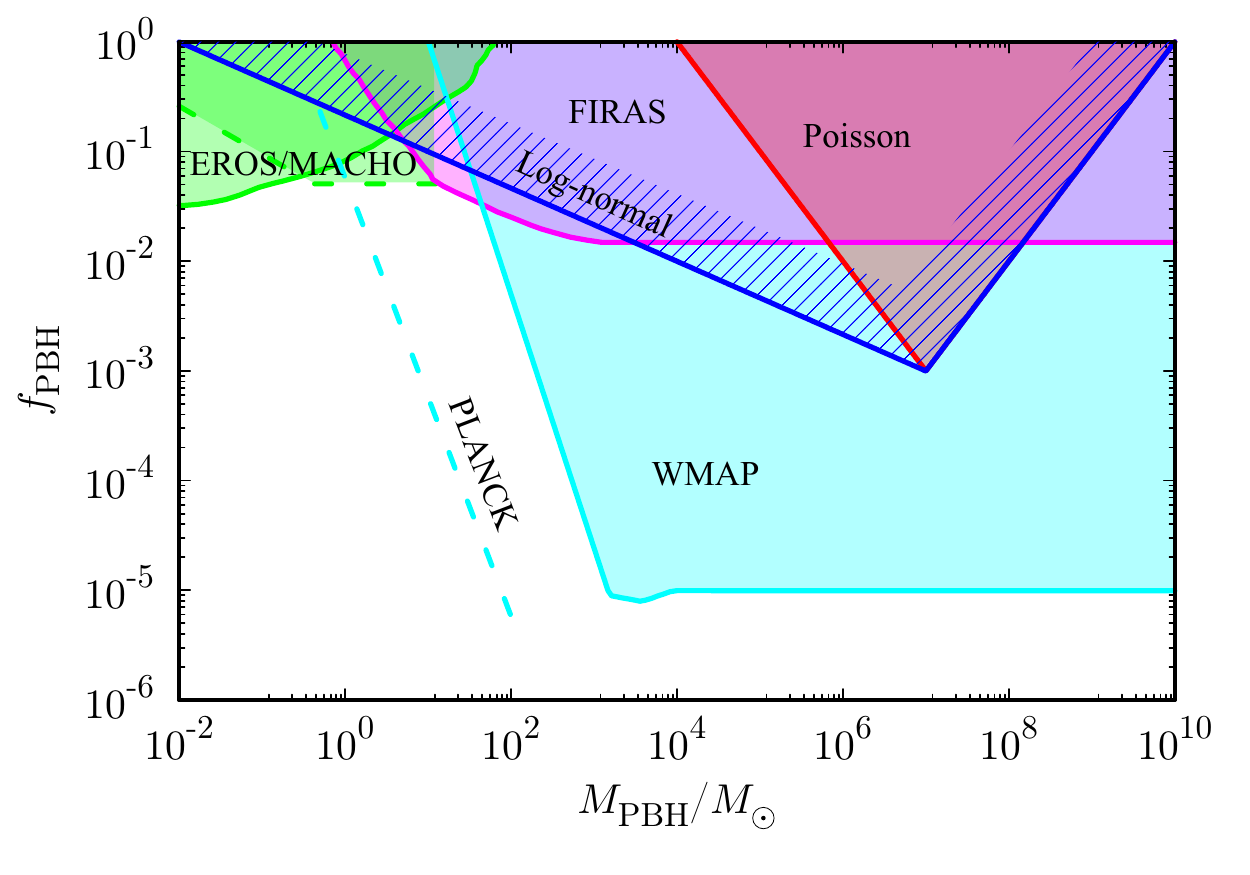}
\caption{
Constraint on the PBH abundance.
The constraint by the Ly-$\alpha$ forest in the case with the log-normal fluctuation corresponds to the blue-hatched region and above.
We have taken $\alpha=1$.
}
\label{fig:PBHconstraint1}
\end{figure}

From the fact that the Poisson fluctuation in the number of PBHs can affect the large-scale structure formation, the PBH abundance can be constrained by the observation of the  Ly-$\alpha$ forest \cite{Afshordi:2003zb}. 
Applying the same discussion to the case with the log-normal fluctuations, we can obtain a stronger constraint on the PBH abundance.
Focusing on the scale corresponding to the Ly-$\alpha$ cloud, we set $\rho_{\rm PBH} V = f_{\rm PBH} M_{{\rm Ly}\alpha}$ with $M_{{\rm Ly}\alpha} \sim 10^{10} M_\odot$. Then, we obtain the variance of the density contrast from \eqref{eq:delta2PBH} and \eqref{eq:NPBH} as
\begin{equation}
\langle \delta^2 \rangle 
= f_{\rm PBH}^2 \left\langle \delta_{\rm PBH}^2 \right\rangle 
\sim 10^{-5} \alpha f_{\rm PBH}^{3/2} \left(\frac{M_{\rm PBH}}{M_\odot} \right)^{1/2} 
\left( \frac{M_{{\rm Ly}\alpha}}{10^{10} M_\odot} \right)^{-1/2} \, .
\end{equation}
Roughly speaking, the Ly-$\alpha$ observation requires that the root-mean-square of the density contrast after the linear growth by a factor of $z_{\rm eq}/z_{{\rm Ly}\alpha} \sim 10^3$ should not exceed $\mathcal{O}(1)$ for the scale of Ly-$\alpha$ cloud \cite{Carr:2009jm}, which yields the constraint 
\begin{equation}
f_{\rm PBH} < 10^{-2/3} \alpha^{-2/3} \left(\frac{M_{\rm PBH}}{M_\odot} \right)^{-1/3}  \left( \frac{M_{{\rm Ly}\alpha}}{10^{10} M_\odot} \right)^{1/3} \, .
\end{equation}
Here we assume that there is at least one PBH in Ly-$\alpha$ cloud, which imposes the following condition:
\begin{equation}
f_{\rm PBH} > \frac{M_{\rm PBH}}{M_{{\rm Ly}\alpha}} \, .
\end{equation}
Figure~\ref{fig:PBHconstraint1} shows the constraint on the PBH abundance in terms of the PBH mass from the above rough estimate.
The observation of the Ly-$\alpha$ forest rules out the blue-hatched region and above in the case with the log-normal fluctuations with $\alpha = 1$.

\subsection{Forecasted constraint by the 21cm forest}

Here we consider the cosmological dark age filled with neutral hydrogens after recombination.
In this epoch, the transition between hyperfine splitting states of the neutral hydrogen atom frequently occurs by emitting and absorbing photons with wavelength 21cm which corresponds to the rest-frame frequency $\nu_* = 1.42$\,GHz. In particular, the emission process is enhanced inside minihaloes and the modification of the halo mass function affects the 21cm emission signals.
In what follows, we follow the analysis in \cite{Iliev:2002gj,Chongchitnan:2012we} (see also \cite{Gong:2017sie}) to calculate the 21cm emission signal from minihaloes with the truncated isothermal sphere as a halo model \cite{Shapiro:1998zp,Iliev:2001he}.

The 21cm signature is characterized by the brightness temperature.
The brightness temperature is obtained as a solution of the radiative transfer equation and the signal we receive is in the form of the differential brightness temperature with respect to the background CMB temperature. Here we focus on the emission signals from a number of minihaloes and in this case, the mean differential brightness temperature over the full sky can be obtained by adding all halo contributions,
\begin{equation}
\overline{\delta T_b} = \frac{c(1+z)^4}{\nu_* H(z)} \int^{M_{\rm max}}_{M_{\rm min}} \Delta \nu_{\rm eff} \delta T_b(M) A \frac{dn}{dM} dM \, ,
\end{equation}
where $\delta T_b(M)$ is the mean value of the differential brightness temperature from single minihalo with halo mass $M$, $\Delta \nu_{\rm eff}$ is the effective redshifted line width, $M_{\rm max}$ is the maximum halo mass determined by the virial temperature $T_{\rm vir} = 10^4$\,K, $M_{\rm min}$ is the minimum halo mass given by the Jeans mass, $A$ is the cross section of single minihalo and $c$ is the speed of light.

Observations can tell us only statistical quantities and the simplest one is the root-mean-square of the fluctuation. The amplitude of the $q$-$\sigma$ fluctuation of the differential brightness temperature over the survey volume is
\begin{equation}
\left\langle \delta T_b^2 \right\rangle^{1/2} = q \sigma_p(\Delta\theta_{\rm beam},\Delta \nu_{\rm band}) \beta(z) \overline{\delta T_b} \, ,
\end{equation}
where $\sigma_p(\Delta\theta_{\rm beam},\Delta \nu_{\rm band})$ is the standard deviation of the linear density perturbation for pencil-beam survey with $\Delta\theta_{\rm beam}$ and $\Delta\nu_{\rm band}$ being the beam angle and the frequency band width respectively and $\beta(z)$ is the flux-weighted average of the bias \cite{Chongchitnan:2012we}. Here we set $q=3$.

\begin{figure}[tp]
\centering
\subfigure[$M_{\rm PBH} = M_\odot$]{
\includegraphics [width = 7.5cm, clip]{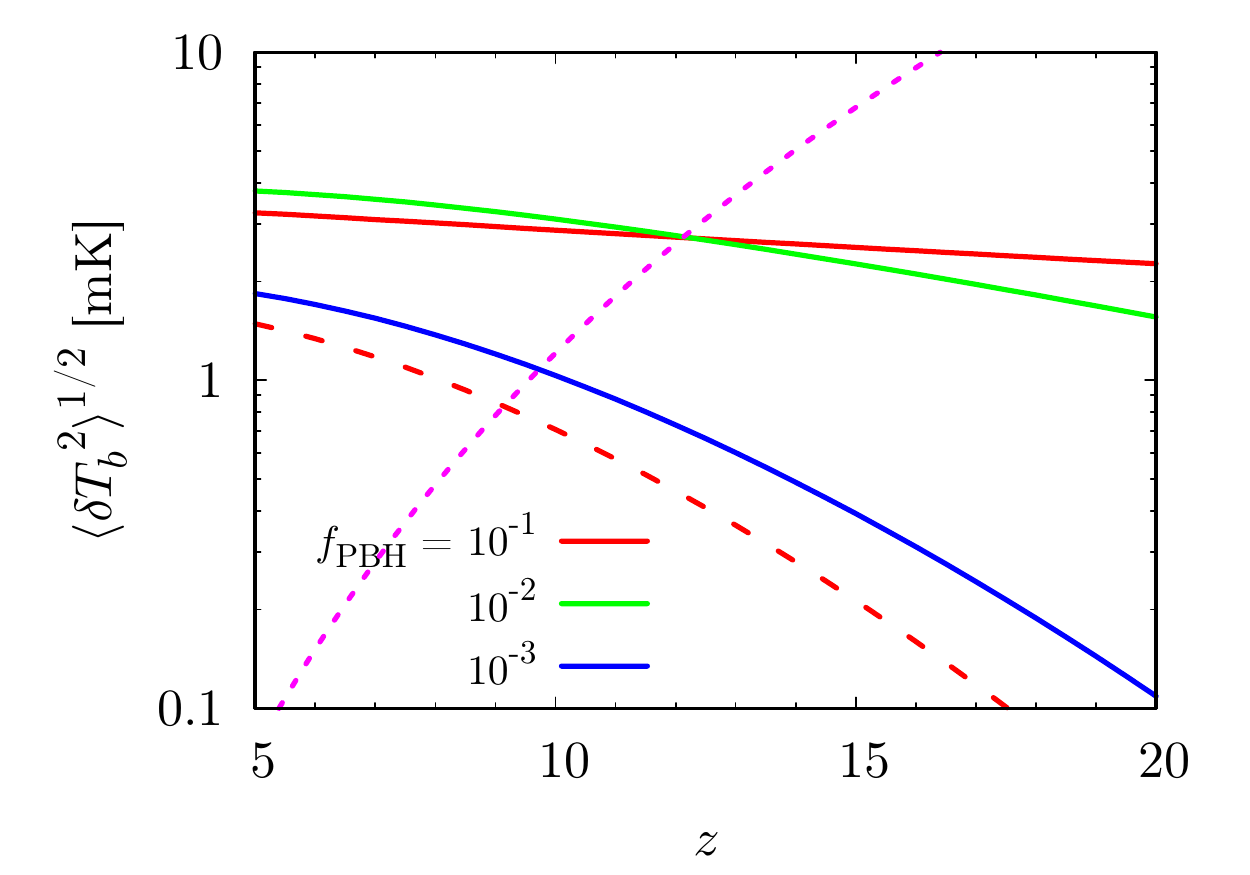}
\label{subfig:dTb1}
}
\subfigure[$M_{\rm PBH} = 10M_\odot$]{
\includegraphics [width = 7.5cm, clip]{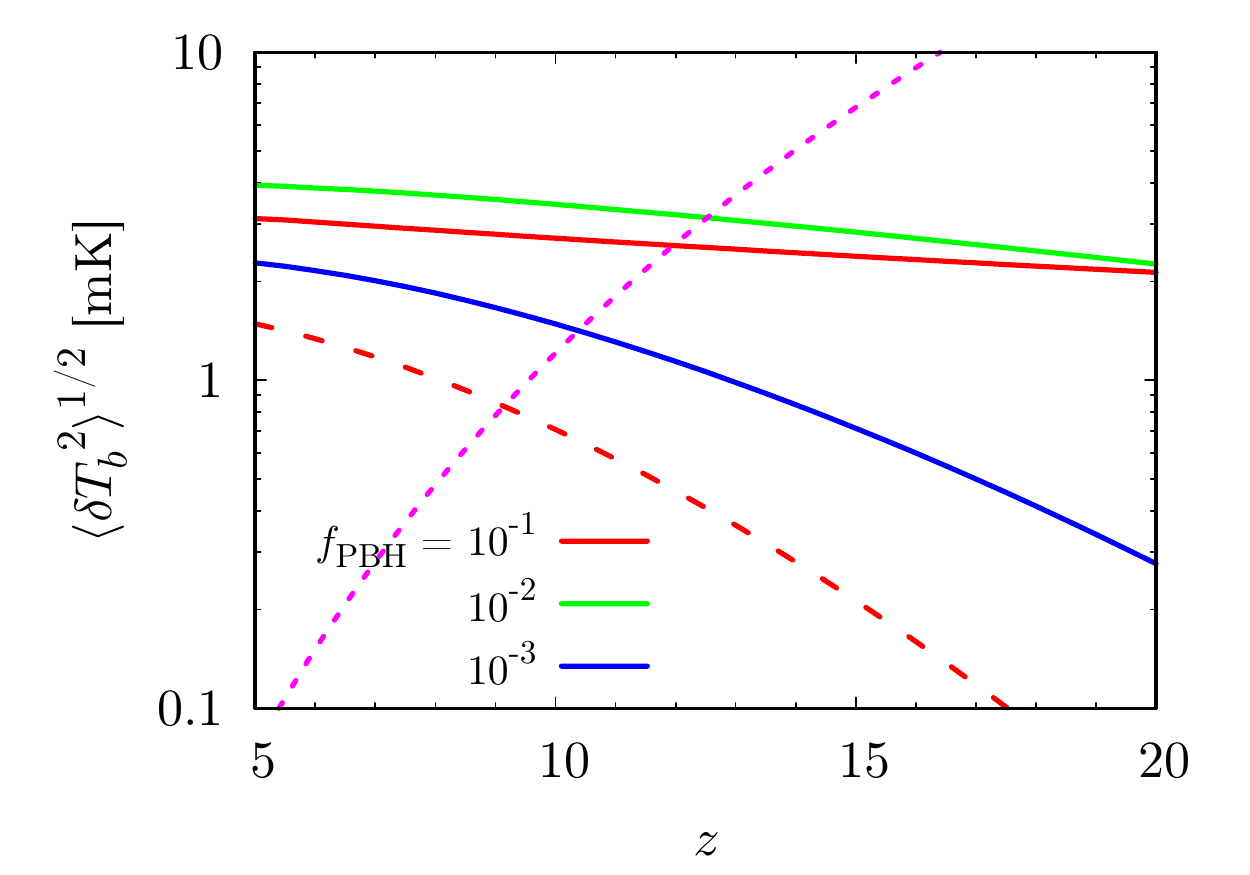}
\label{subfig:dTb2}
}
\subfigure[$M_{\rm PBH} = 100M_\odot$]{
\includegraphics [width = 7.5cm, clip]{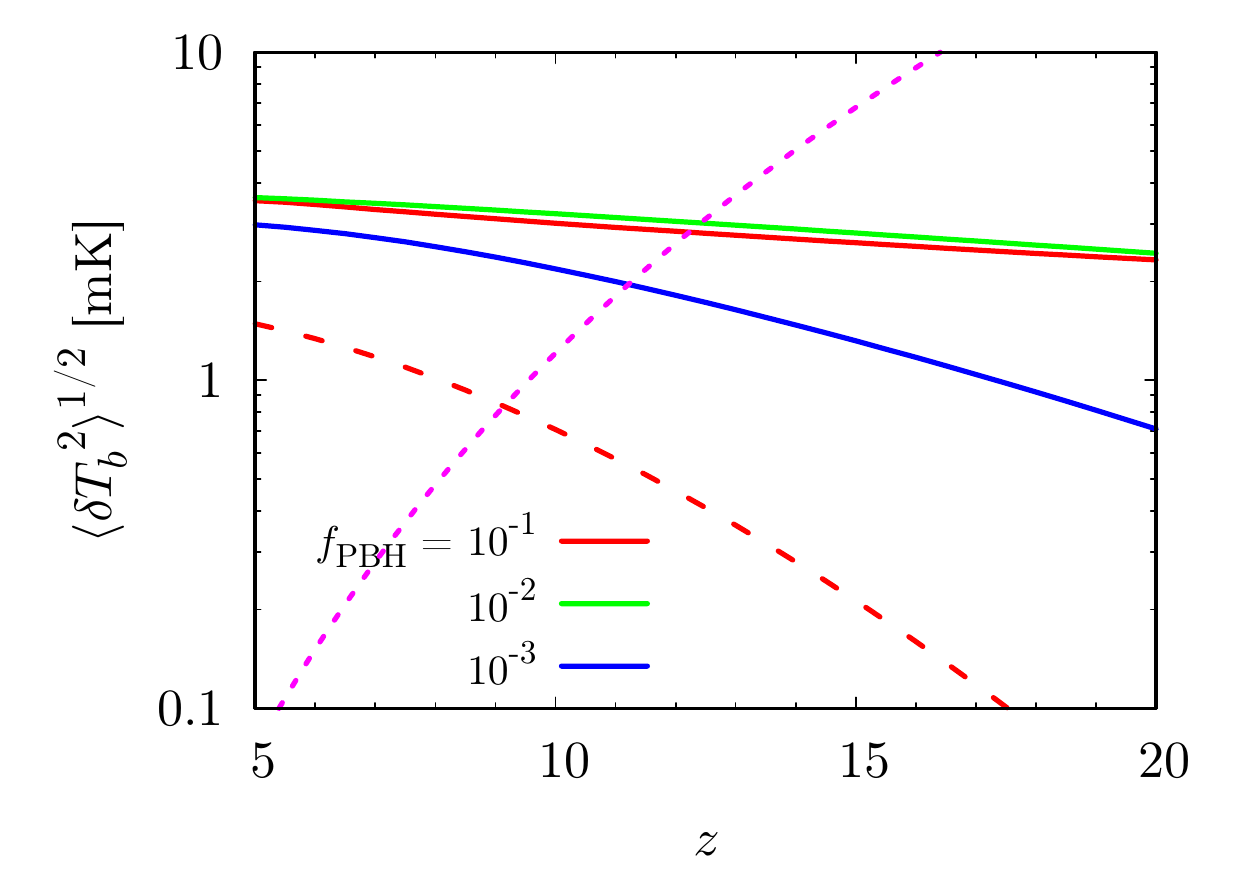}
\label{subfig:dTb3}
}
\subfigure[$M_{\rm PBH} = 1000M_\odot$]{
\includegraphics [width = 7.5cm, clip]{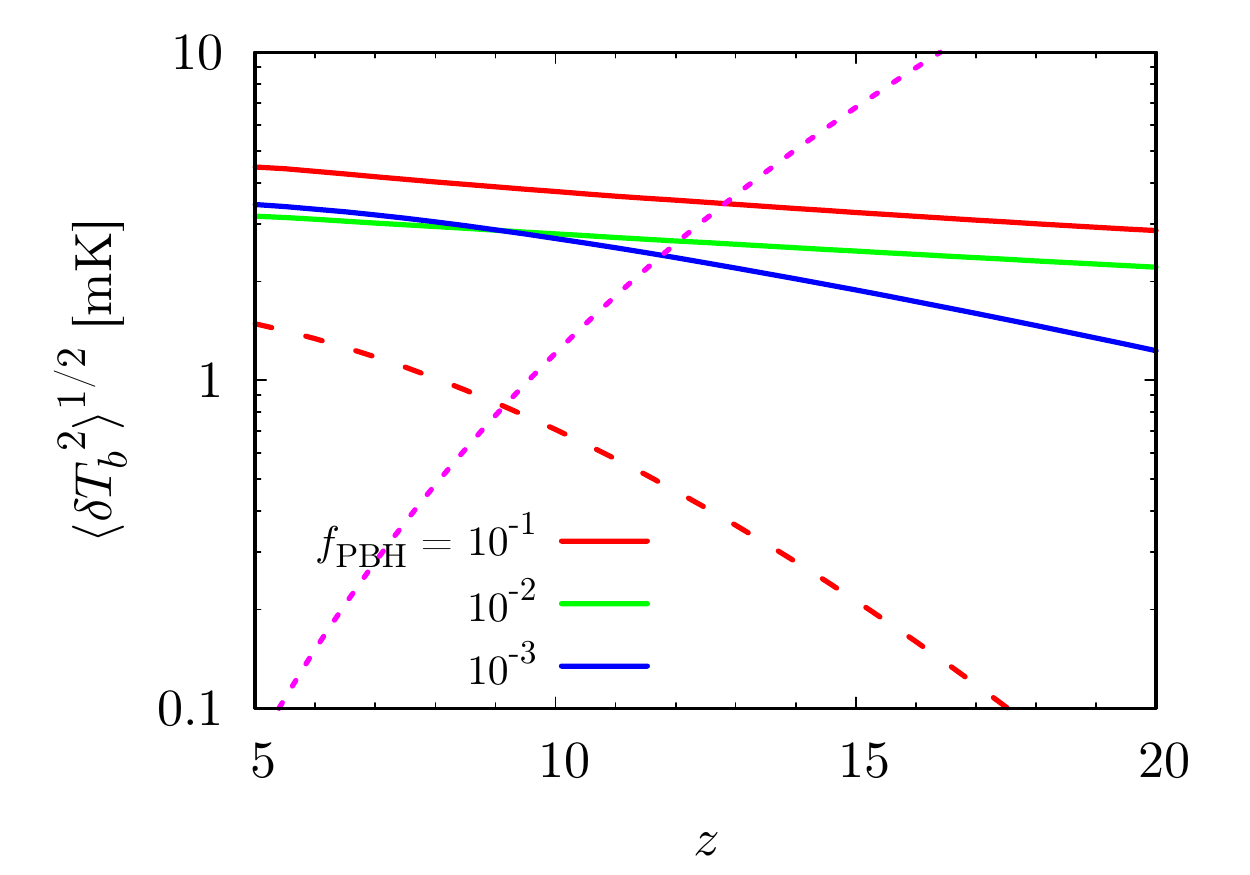}
\label{subfig:dTb4}
}
\caption{
The root-mean-square value of the brightness temperature fluctuation. We have taken $\alpha=1$, $z=10$, $M_{\rm PBH} = M_\odot$ (a), $10M_\odot$ (b), $M_{\rm PBH} = 100M_\odot$ (c), $1000M_\odot$ (d) and $f_{\rm PBH} = 10^{-1}$ (red), $10^{-2}$ (green) and $10^{-3}$ (blue). The dashed red line shows the case without PBH and the dotted magenta line shows the noise curve of SKA.
}
\label{fig:dTb}
\end{figure}

The noise of the SKA-like observation is \cite{Furlanetto:2006jb}
\begin{equation}
\delta T_{\rm noise} =  20\,{\rm mK}\,\frac{10^4\,{\rm m}^2}{A_{\rm tot}} \left( \frac{10\,{\rm arcmin}}{\Delta\theta_{\rm beam}} \right)^2 \left(\frac{1+z}{10} \right)^{4.6} \left( \frac{\rm MHz}{\Delta \nu_{\rm band}} \frac{100\,{\rm h}}{t_{\rm int}} \right)^{1/2} \, ,
\end{equation}
with $A_{\rm tot}$ and $t_{\rm int}$ being the effective collecting area of the radio telescope arrays and the integration time respectively. Here we set $A_{\rm tot}=10^5\,{\rm m}^2$, $\Delta\theta_{\rm beam}=9$\,arcmin, $\Delta\nu_{\rm band}=1$\,MHz and $t_{\rm int}=1000$\,h. Figure \ref{fig:dTb} shows the root-mean-square of the differential brightness temperature as a function of the redshift. Non-trivial dependence on $f_{\rm PBH}$ reflects the non-trivial feature of the mass function.

\begin{figure}[tp]
\centering
\includegraphics [width = 12cm, clip]{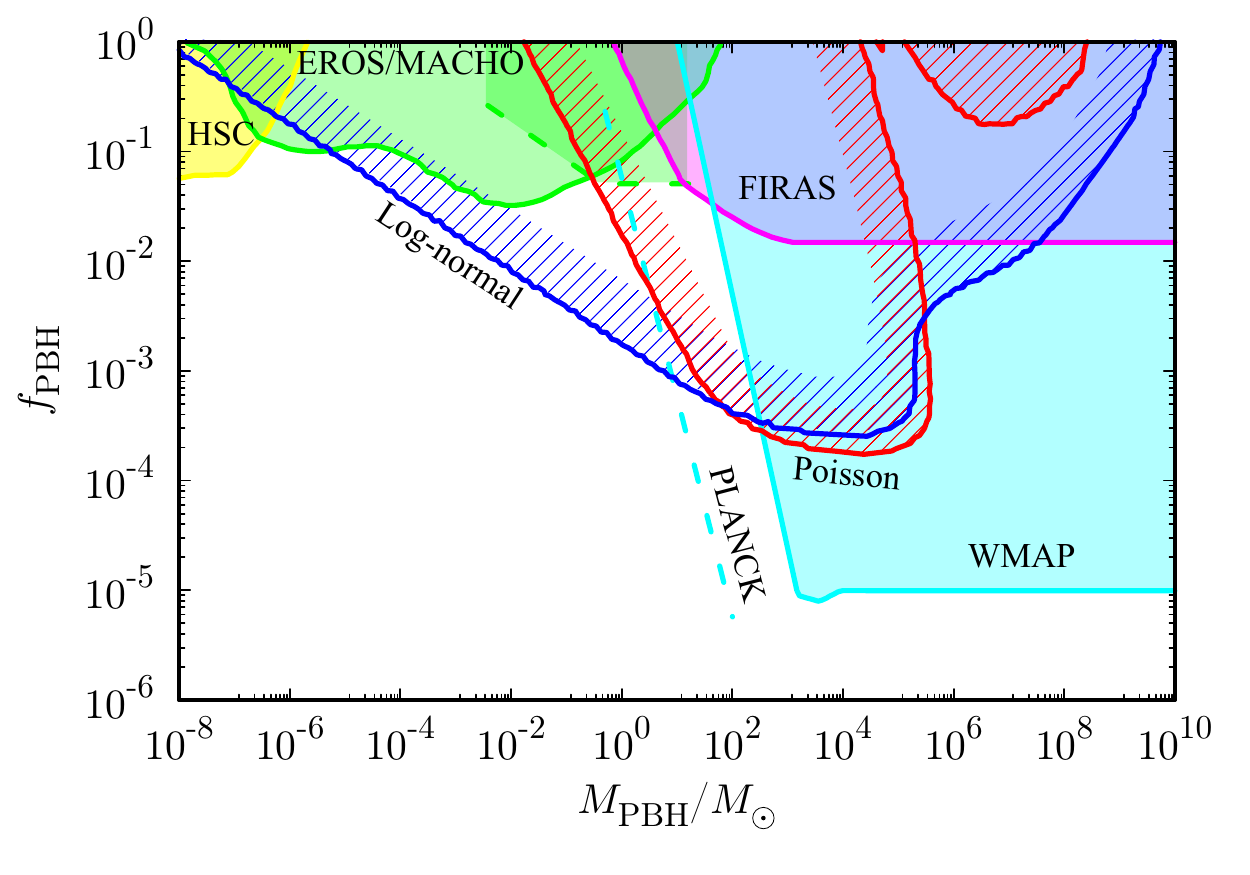}
\caption{
Forecasted constraint on the PBH abundance by future SKA-like observations.
The blue hatched and red hatched areas correspond to the log-normal and the Poisson fluctuations respectively.
}
\label{fig:PBHconstraint2}
\end{figure}

To roughly estimate a forecasted constraint on the PBH abundance by SKA-like observations, we have simply performed $\Delta \chi^2$ analysis following \cite{Sekiguchi:2013lma}. The result is shown in Figure~\ref{fig:PBHconstraint2}.
The log-normal fluctuation on the number of PBHs can predict the 21cm signature with detectable level for $10^{-5}M_\odot \lesssim M_{\rm PBH} \lesssim 10M_\odot$, competing with the microlensing constraint, even if the PBH abundance is as small as sub-percent of the total dark matter abundance.

\section{Discussion} 
\label{sec:disc}

In this article, we have considered the distribution of the PBH number fluctuation in the case where the primordial perturbation with a monochromatic power spectrum seeds the PBH formation.  We have shown that for a steeply blue-tilted spectrum with $n_\delta \gtrsim 3$, the distribution of the PBH number follows the Poisson distribution but for moderately tilted case with $n_\delta \lesssim 2$, the distribution follows the log-normal distribution. The log-normal distribution enhances the isocurvature fluctuations on even larger scales compared with the Poisson case and the halo mass function is modified significantly. The constraint on the PBH abundance from the Ly-$\alpha$ observation is modified accordingly and the future observations of the 21cm fluctuation can potentially put strong constraints over a wide range of PBH masses, $10^{-5} M_\odot < M_{\rm PBH} < 10M_\odot$.
We have made only a rough estimate for the above observables. More detailed discussion including the physics during the reionization and forecasted constraints making use of the Fisher analysis are beyond the scope of this article and left for future work.

Throughout this article, we have considered only the power spectrum of the density contrast with a constant spectral index and have not discussed the connection with the primordial curvature perturbation and specific inflation models.
In fact, to realize the mildly blue-tilted spectrum for density contrast, we need highly red-tilted power spectrum of curvature perturbation. In that case, however, heavier PBHs are more abundantly produced at later time, leading to the overproduction. One of the possibility to evade such a problem is the curvaton scenario where the super-horizon scale curvature perturbation evolves dynamically.
More detailed analysis on the distribution of PBHs and its small-scale signature given the primordial curvature perturbation with specific inflation and curvaton models including the running of the spectral index is left for future work.

\section*{Acknowledgments}

We thank
Bernard Carr, Misao Sasaki, Takahiro Tanaka, Atsushi Taruya, Shuichiro Yokoyama and Chul-Moon Yoo
for helpful discussions.
We are grateful to the Yukawa Institute for Theoretical Physics at Kyoto University for hospitality during the workshop YITP-T-17-02 ``Gravity and Cosmology 2018'' and the YKIS symposium YKIS2018a ``General Relativity -- The Next Generation --'' while this work was under progress. 
JG is supported in part by the Basic Science Research Program through the National Research Foundation of Korea Research Grant 2016R1D1A1B03930408, and by a TJ Park Science Fellowship of POSCO TJ Park Foundation.
NK acknowledges the support by Grant-in-Aid for JSPS Fellows.

\end{document}